# Small-angle physics at the intersecting storage rings forty years later

*Ugo Amaldi*

University Milano Bicocca and TERA Foundation

## 1  Hadron–hadron cross-sections at the beginning of the 1970s

The first surprising 'small-angle physics' result produced at the Intersecting Storage Rings (ISR) was the announcement that the proton–proton total cross-section was not constant over the newly opened energy range. Since the present report is not a review paper but a personal recollection on the beginnings of ISR physics, I start by underlining that, forty years later, it is difficult to describe and explain the surprise and scepticism with which the news of the 'rising total cross-section' was received by all knowledgeable physicists. Among the many episodes, I vividly recall what Daniele Amati told me while walking out of the CERN Auditorium after the seminar of March 1973 in which I had described the results obtained independently by the CERN–Rome and Pisa–Stony Brook Collaborations: 'Ugo, you must be wrong, otherwise the pomeron trajectory would have to cut the axis above 1!'

Nowadays, all those who still care about the pomeron know the phenomenon, find it normal and accept the explanations of this fact given by the experts. But at that time the reggeon description of all small-angle hadronic phenomena was the only accepted dogma since it could explain the main experimental results in hadronic physics: (i) the tendency of the total cross-sections of all hadron–hadron collisions to become energy independent; and (ii) the 'shrinking' of the forward differential cross-sections when the collision energy was increasing.

The second phenomenon was discovered in 1962 by Bert Diddens, Alan Wetherell *et al.*, who were studying proton–proton collisions at the CERN Proton Synchrotron (PS) [1]. They found that the forward proton–proton differential elastic cross-section at small centre-of-mass angles $\theta_{cm}$ (i.e., at small momentum transfers $q = cp_{cm} \sin\theta_{cm}$ usually measured in GeV) is proportional to $\exp(-Bq^2)$, with a slope parameter $B$ that *increases* with the centre-of-mass energy, indicating, through the uncertainty principle, that the proton–proton interaction radius *increases* as $\sqrt{B}$.

As far as the first point is concerned, it was later said that the results shown in Fig. 1 – which had been obtained in the early 1970s at the Protvino 70 GeV synchrotron [2] – were already showing that the Regge model could not describe the rise of the $K^+$–proton cross-section with energy.

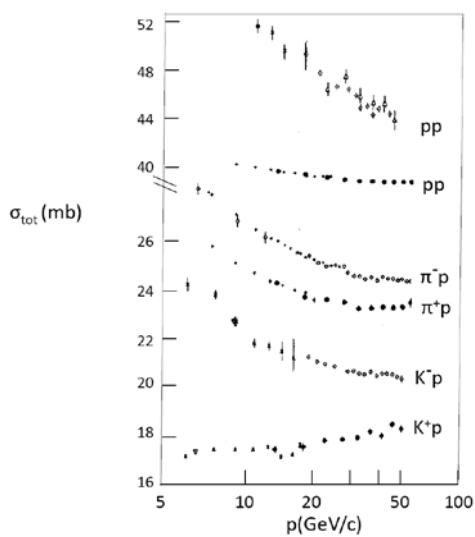

**Fig. 1:** The total cross-sections $\sigma_{tot}$ (measured in the early 1970s at the Serpukhov 70 GeV synchrotron and at lower-energy accelerators) plotted versus the laboratory proton momentum $p$ [2]

However, these experimental results were immediately interpreted in the framework of the Regge model, and even the authors did not conclude that there was an indication of an anomaly in the energy dependence of the cross-section of this particular channel. With reference to Fig. 1, in the paper by Denisov *et al.* [2], which was also signed by Jim Allaby and Giorgio Giacomelli and was received by *Physics Letters* in July 1971 – three months after the ISR start-up – one can read: 'This figure suggests that the total cross-section for $K^+$–p will approach the asymptotic value from below … unless the cross-section oscillates in value.'

The regime in which all the total cross-sections would become energy independent was called 'asymptopia', and theorists and experimentalists alike were convinced that the ISR would demonstrate that the total proton–proton cross-section, which slightly decreases in the Serpukhov energy range (Fig. 1), would tend to a constant of about $40 \times 10^{-26}$ cm$^2$ (40 mb), thus confirming the mainstream interpretation of all hadronic phenomena, the Regge model, to be discussed in the next section.

## 2 The theoretical framework

As far as the hadronic forward differential cross-sections were concerned, they had found a universally accepted interpretation in terms of the collective effect of the exchanges of all the particles, which, in the mass$^2$–spin plane, lay on a Regge 'trajectory'. In Fig. 2, the present knowledge of the ρ trajectory is reported [3].

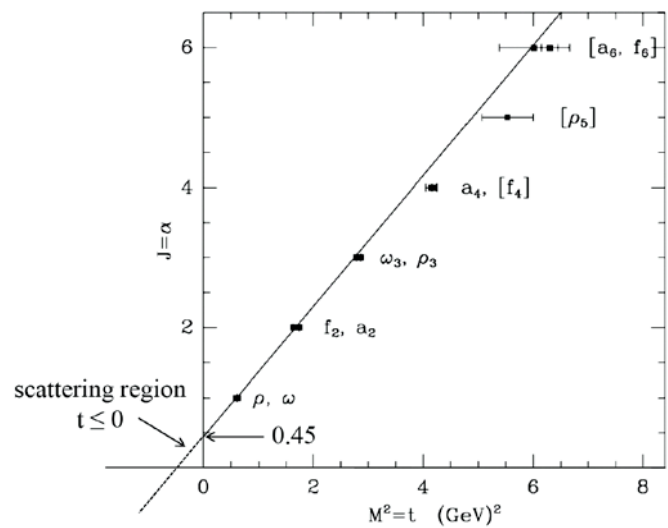

**Fig. 2:** The present situation of the Chew–Frautschi plot shows that the Regge trajectory containing the ρ meson (mass = 770 MeV) is practically linear up to very large masses

The exchange of the ρ trajectory dominates the charge-exchange cross-section of Fig. 3a. By using the usual parameter $s = E_{cm}^2$, where $E_{cm}$ is the centre-of-mass energy, the recipes of the Regge model give a cross-section that varies as $s^{\alpha(t=0)-1}$.

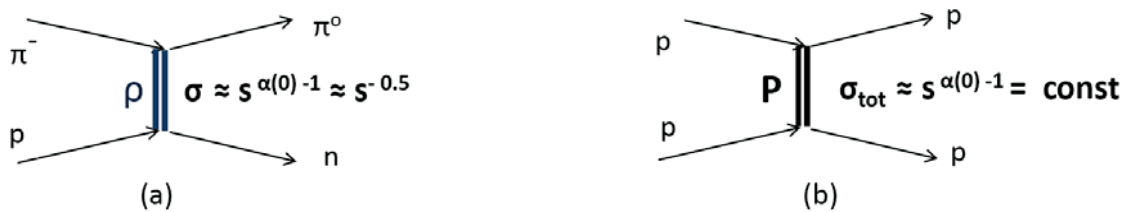

**Fig. 3:** (a) The main contribution to the pion charge-exchange phenomenon is the exchange of the ρ trajectory. (b) In the Regge model, the exchange of a pomeron trajectory is the dominant phenomenon in all high-energy elastic collisions.

Since, in Fig. 3a, $\alpha(0) \approx 0.5$, the charge-exchange cross-section was predicted to vary roughly as $s^{-0.5} = 1/E_{cm}$. In the 1960s the experimental confirmation of this prediction was one of the strongest arguments in favour of the Regge description of the scattering of two hadrons. Such a description is still used because these phenomena cannot be computed with quantum chromodynamics – the strong interacting theory of the fundamental components of all hadrons.

A second argument concerned the differential cross-sections $d\sigma/d|t|$, which defines the behaviour of the trajectory in the region of the plot in Fig. 2 indicated as 'scattering region', where the variable $m^2$ becomes $t = -q^2$.

At the time that the rising cross-section was reported, accurate preliminary data on the charge-exchange differential cross-section were coming once again from Yuri Prokoshkin's group working at the Serpukhov accelerator [4]. Figure 4 shows how well the data in the 'scattering region' of Fig. 2 join the slope value of the ρ trajectory in the positive $m^2$ region.

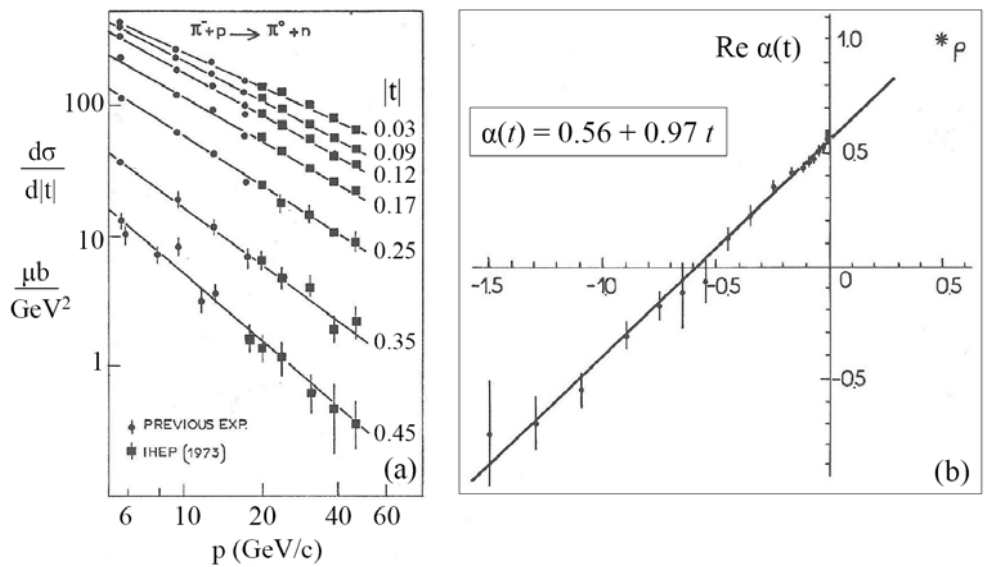

**Fig. 4:** (a) The exchange of a linear ρ trajectory fits the experimental data very well [4]. (b) In this experiment, the trajectory was found to be linear in the range $-1.5 \leq t \leq 0$ GeV$^2$ and the derivative of $\alpha(t)$ was measured to be $\alpha'(0) = 1$ GeV$^{-2}$ at $t = 0$.

As shown in Fig. 3b, in the Regge approach, the proton–proton scattering process was also described by the exchange of a trajectory, the pomeron, which, given the proportionality of $\sigma_{tot}$ to $s^{\alpha(t=0)-1}$, had to have the value $\alpha_p(t=0) = 1$ to be consistent with an energy-independent total cross-section. For this reason, at the beginning of the 1970s, the so often heard 'asymptopia' and 'the pomeron intercept is equal to 1' were used as different ways of saying the same thing.

Since there were no particles belonging to the pomeron trajectory, its slope could be fixed only by measuring the $t$ dependence of the forward elastic proton–proton cross-section, as done for the exchange cross-section with the data of Fig. 4. Here we meet the already quoted argument in favour of the reggeon model: the forward proton–proton elastic cross-section could be described by the simple exponential $\exp^{-B|t|}$ and that the 'slope' $B$ *increases* with the centre-of-mass energy. This is described by saying that 'the forward peak shrinks with energy', a statement that we now know applies to most high-energy differential cross-sections.

The shrinking of the forward peak measured at Serpukhov confirmed earlier data and indicated that the slope of the pomeron trajectory is about three times smaller than that of the ρ trajectory, which is about 1 GeV$^{-2}$ (Fig. 4b). At the time the ISR was constructed, the determination of the slope $B$ – easy to measure in a new and large energy range – was considered a very important issue.

In parallel with this 't-channel' description, other theorists, working on the 's-channel description', were deriving rigorous mathematical consequences from the fundamental properties of the S-matrix, which describes the scattering processes: unitarity, analyticity and crossing. Unitarity of the S-matrix implies that one can compute the imaginary part of the forward scattering amplitude Im $f(t)$ by taking the product of a scattering amplitude and its conjugate and summing them over all possible intermediate states, as graphically depicted in Fig. 5.

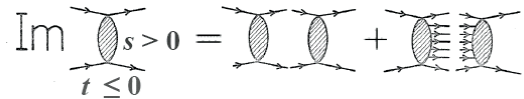

$$4\pi \, \text{Im} \, f(t)/k = G_{\text{el}}(t) + G_{\text{in}}(t).$$

**For $t = 0$:** $4\pi \, \text{Im} \, f(0)/k = \sigma_{el}(s) + \sigma_{inel}(s) = \sigma_{tot}(s)$

**Fig. 5:** The graphical representation of the unitarity relation, at a given $s$ and for $t \leq 0$, explains the definition of the elastic and inelastic overlap integrals $G_{\text{el}}(t)$ and $G_{\text{in}}(t)$. In the equations, $k = p/\hbar$.

The sum is made up of two contributions, which are called 'elastic and inelastic overlap integrals' $G_{\text{el}}(t)$ and $G_{\text{in}}(t)$. In the forward direction, i.e., for $t = 0$, the overlap integrals reduce to the elastic and inelastic cross-sections, and the unitarity relation gives the 'optical theorem', which states that the imaginary part of the forward scattering amplitude equals the total cross-section $\sigma_{\text{tot}}$, except for a factor $4\pi/k$, which depends on the definition chosen for the amplitude itself.

The figure and the formulae indicate that hadron–hadron forward elastic scattering is determined by the amplitudes of both elastic and inelastic reactions. When the collision energy is large, there are many opened inelastic channels, the incoming wave is absorbed and the elastic scattering amplitude is dominated by its imaginary part, which is the 'shadow' of the elastic and inelastic processes. In such a *diffraction phenomenon*, the ratio $\rho = \text{Re}(f)/\text{Im}(f)$ between the real and imaginary parts of the elastic amplitude is expected to be small, so that, in the expression for the forward elastic cross-section deduced from the optical theorem,

$$\left(\frac{d\sigma_{\text{el}}}{d|t|}\right)_{t=0} = \frac{(1+\rho^2)\sigma_{\text{tot}}^2}{16\pi} \quad \text{with} \quad \rho = \frac{\text{Re} \, f(0)}{\text{Im} \, f(0)},$$

the term $\rho^2$ is of the order of a few per cent.

Combining unitarity with analyticity and crossing, in the 1960s three important theorems had been demonstrated.

- The *Pomeranchuk theorem* [5] states that, in the limit $s \to \infty$, the hadron–hadron and the antihadron–hadron cross-sections become equal.

- According to the *Froissart–Martin theorem* [6, 7] the total cross-section should satisfy the bound

$$\sigma_{\text{tot}} \leq C \ln^2(s/s_0) \approx 60 \text{ mb } \ln^2(s/s_0)$$

  where the numerical value $C = \pi(\hbar/m_\pi)^2$ is determined by the mass of the pion, which is the lightest particle that can be exchanged between the two colliding hadrons, and $s_0$ is usually taken equal to 1 GeV$^2$.

- Finally, the *Khuri–Kinoshita theorem* [8] relates the energy dependence of $\rho$ with the energy dependence of the total cross-section by stating that, if $\sigma_{\text{tot}}$ increases with energy, $\rho$ passes from small negative values to positive values. This is a consequence of the 'dispersion relations', which connect the real part of the forward elastic amplitude with some appropriate energy integrals of the total cross-section. Khuri and Kinoshita showed that, if $\sigma_{\text{tot}}$ follows the Froissart–Martin bound and increases proportionally to $\ln^2 s$, for $s \to \infty$, the ratio $\rho$ is *positive* and tends to zero from above towards the horizontal axis proportionally to $\pi/\ln s$.

In summary, the reggeon (*t*-channel) description of hadron–hadron collisions and the theorems derived from the properties of the *S*-matrix were the theoretical tools available at the end of the 1960s to experimental physicists interested in total cross-sections and small-angle physics.

## 3 Three proposals

In March 1969 the ISR Committee received three proposals that are relevant to the subjects discussed in this paper.

The title of the proposal by the Pisa group (signed by G. Bellettini, P.L. Braccini, R.R. Castaldi, C. Cerri, T. Del Prete, L. Foà, A. Menzione and G. Sanguinetti) was 'Measurements of the p–p total cross section' [9]. Two of their figures are reproduced in Fig. 6. The very large scintillator hodoscopes would detect the outgoing particles and count the total number of events. Moreover, the small-angle telescope, not shown in the figure, would detect forward elastic events and extrapolate to zero scattering angle the differential cross-section to estimate the number of elastic events not recorded because the protons, scattered at small angles, would be lost in the ISR vacuum chamber.

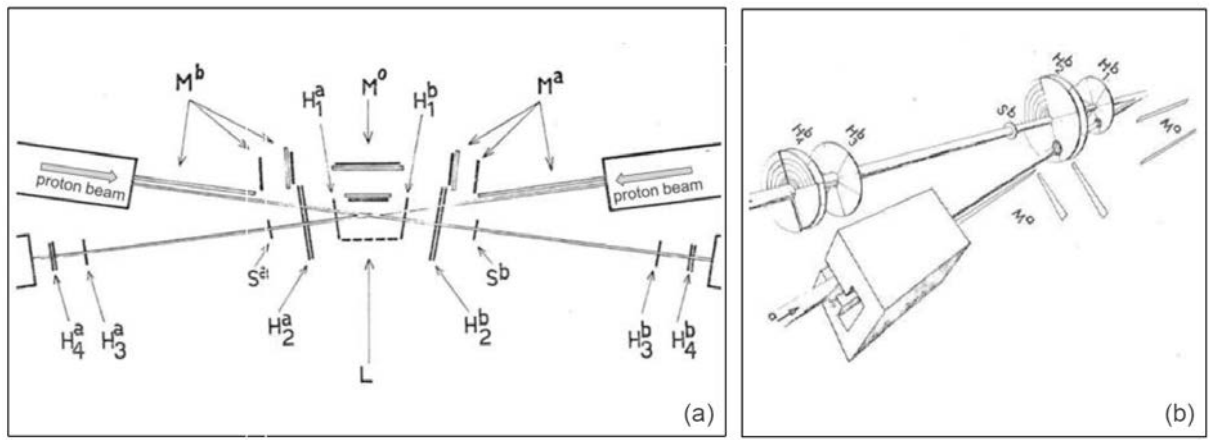

**Fig. 6:** The initial proposal by the Pisa group to measure the proton–proton total cross-section

To compute any cross-section, one needs a measurement of the 'luminosity' $L$. In the case of a beam of parallel particles that cross at an angle, the only important spatial variable is the vertical one $y$. Given the normalized vertical distributions of the two beams, $\rho_1(y - y_o)$ and $\rho_2(y)$, which are displaced vertically by $y_o$, the luminosity is proportional to the two currents and depends upon the crossing angle of the beams according to the formula:

$$L(y_o) = \underbrace{\frac{I_1 I_2}{c\,e^2\,\tan(\phi/2)}} \underbrace{\int \rho_1(y - y_o)\, \rho_2(y)\, dy}$$
$$R(y_o) = \sigma \cdot K \cdot \text{(overlap integral)}$$

To obtain the luminosity, the Pisa group proposed to measure $\rho_1$ and $\rho_2$ separately, with the two sets of spark chambers indicated in Fig. 6 with the letters M$^o$ and M$^a$, and then to compute the beam overlap integral numerically.

The problem of measuring the ISR luminosity was amply debated during 1968 and various proposals to do so by *separated measurements* of the vertical distributions were put forward by Darriulat and Rubbia [10], Rubbia [11], Schnell [12], Steinberger [13] and Onuchin [14].

Another method proposed in different forms by Cocconi [15], Di Lella [16] and Rubbia and Darriulat [17] was based on the detection of the two protons scattered at angles smaller than about 1 mrad, where the *known* Coulomb elastic scattering cross-section dominates.

All the proposals requiring the separate measurements of the vertical distributions of the two beams were superseded by a very simple observation made by Simon Van der Meer [18]. He remarked that the cross-section $\sigma_M$ of a particular type of event (detected by a set of monitor counters surrounding the interaction region) can be obtained by measuring the rate of the monitor events $R_M(y_o)$ as a function of the distance $y_o$ between the centres of the two beams, which are moved vertically in small and precisely known steps.

In the integral $I_{VdM} = \int R_M(y_o)\, dy_o$ the double integral over $dy_o$ and $dy$ equals 1, because $\rho_1$ and $\rho_2$ are normalized, the cross-section of the monitor counters is given by $\sigma_M = I_{VdM}/K$ and the cross-section $\sigma$ corresponding to any other rate $R$ is simply obtained as

$$\sigma = \frac{R}{R_M}\sigma_M = \frac{I_{VdM}}{K}\frac{R}{R_M}$$

The magnets needed to precisely displace the two beams vertically were installed in the ISR, and since then the Van der Meer method has been used to measure proton–proton luminosities. A typical distribution obtained at the ISR is shown in Fig. 7 [19].

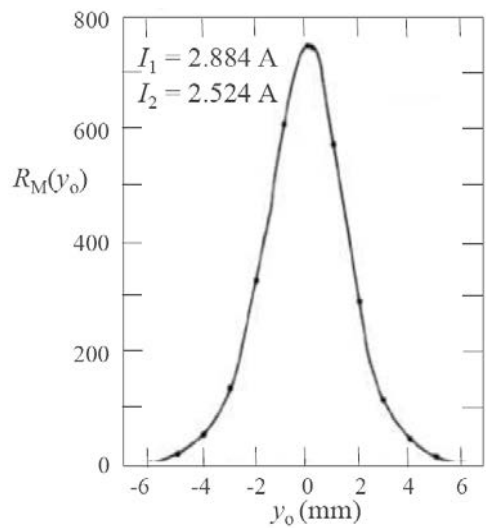

**Fig. 7:** Distribution of a monitor rate versus the vertical distances between two ISR beams

Figure 8 shows the apparatus built by what became the Pisa–Stony Brook Collaboration after joining with the Stony Brook Group led by Guido Finocchiaro and Paul Grannis.

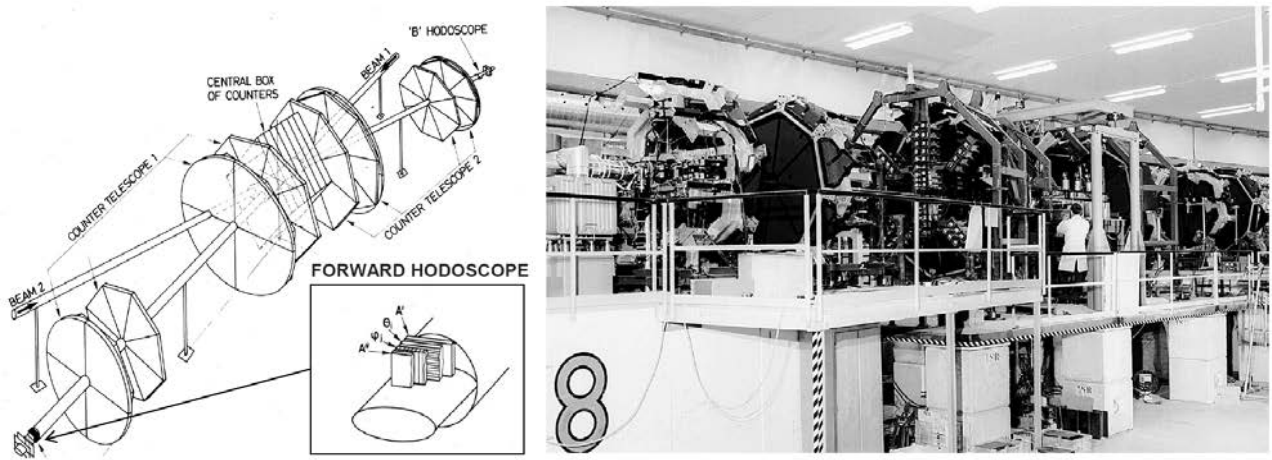

**Fig. 8:** In the final detector built by the Pisa-Stony Brook Collaboration, forward telescopes were used to measure elastic scattering events at small angles

Coulomb scattering was the focus of the proposal 'The measurement of proton–proton differential cross-section in the angular region of Coulomb scattering at the ISR' [20] by the Rome–Sanità group and Paolo Strolin, who at the time was an ISR engineer (signed by U. Amaldi, R. Biancastelli, C. Bosio, G. Matthiae and P. Strolin). The apparatus (shown in Fig. 9) required a modification of the ISR vacuum pipe, and two quadrupoles and one bending magnet had to be installed on each beam.

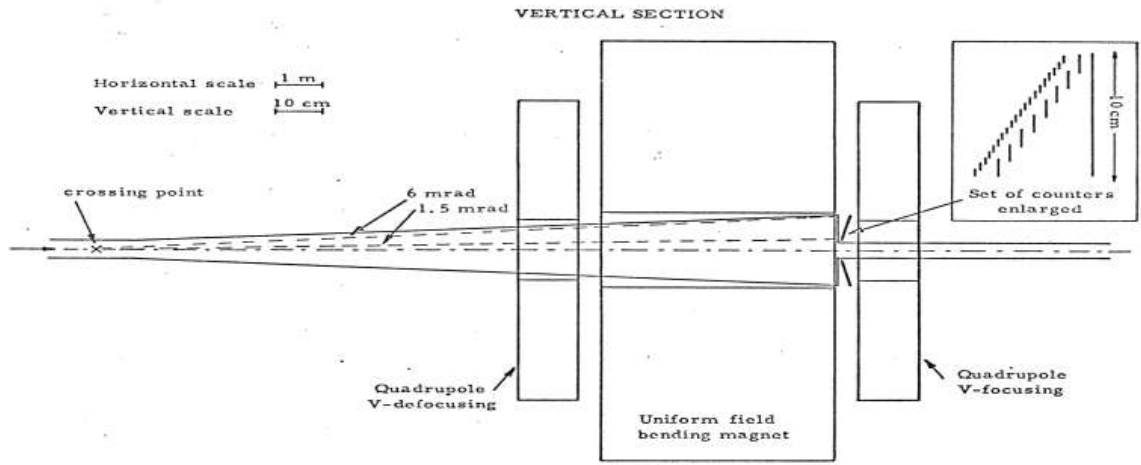

**Fig. 9:** In the first proposal, two quadrupoles and one magnet focused the protons and bent them so as to measure protons scattered down to 1.5 mrad

A few months later, in an addendum to the proposal, the authors wrote: 'In discussions with the specialists of the machine (R. Calder and E. Fischer) we found a simple way for allocating the detectors near the beam, which does not imply a modification of the standard parts of the vacuum chamber.'

The proposal (Fig. 10) foresaw getting as close as 10 mm to the beam with the bottom of the movable sections, as proposed many years before by Larry Jones [21]. This was a daring operation and many people worried so much that, in an ISR meeting, Carlo Rubbia said: 'Your scintillators will give light as bulbs!'

To counter the criticisms, in 1970 a test was performed at the CERN PS to check whether one could install scintillation counters very close to a circulating proton beam. Previously Hyams and Agoritsas had performed similar measurements [22].

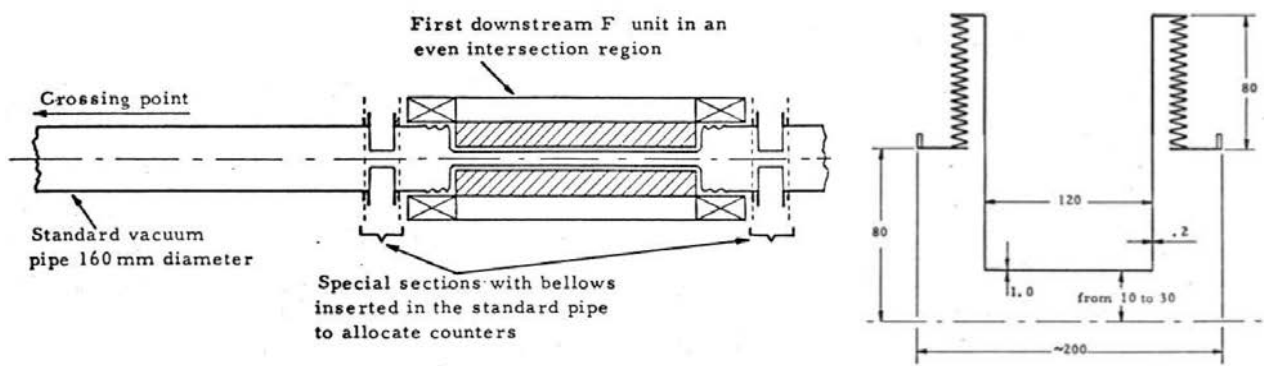

**Fig. 10:** In the 1969 proposal there were four movable sections on each beam and the forward-scattered protons were detected by a coincidence between counters located upstream and downstream of the first ISR magnet

Eifion Jones participated in the planning and in the tests – in which the PS beam was moved towards the scintillators – and a memorandum was sent to the ISR Committee [23], which concluded that, down to a few millimetres from the beam, the rate to be found at the ISR would have been sufficiently low to allow the Coulomb experiment (Fig. 11).

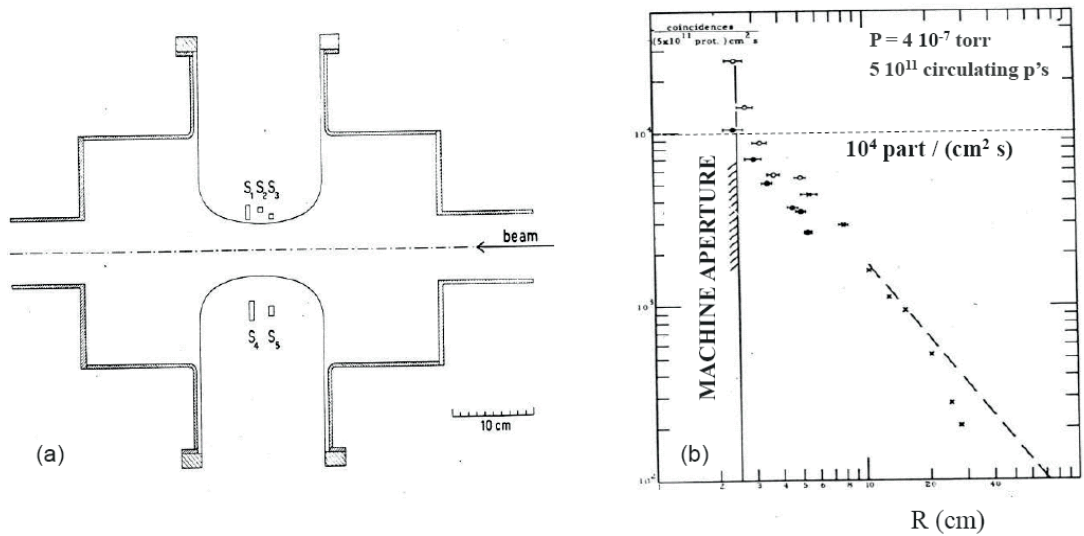

**Fig. 11:** Special section of the PS that allowed the measurement of the rate detected by scintillators placed very close to a circulating beam formed by $5 \times 10^{11}$ protons

The ISR movable sections of the vacuum chamber soon became known as 'Roman pots', which was the translation of the expression '*les pots de Rome*' invented by the French draftsman whom we visited regularly travelling from Rome to Geneva and who, under the direction of Franco Bonaudi, transformed our rough sketches into construction drawings.

In October 1970 the ISR Committee took various decisions on pending experiments. Following it, the CERN group of Giuseppe Cocconi, Alan Wetherell, Bert Diddens and Jim Allaby wrote the Committee a memo, which said: 'At the meeting of the ISRC on 14 October it was concluded that there is no way to fit the proposed experiment on deep inelastic scattering into the present ISR experimental program. As a result we have decided, on their invitation, to collaborate with the Rome group (U. Amaldi *et al.*) on the small-angle scattering experiment.'

For the final experiment, the newly formed CERN–Rome Collaboration decided to retain only the four movable sections located in front of the first ISR magnet, a decision that simplified the experiment and its interactions with the accelerator.

The title of the proposal by the CERN–Genoa–Torino group (P. Darriulat, C. Rubbia, P. Strolin, K. Tittel, G. Diambrini, I. Giannini, P. Ottonello, A Santroni, G. Sette, V. Bisi, A Germak, C. Grosso and M.I. Ferrero) was 'Measurement of the elastic scattering cross-section at the ISR' [24]. The apparatus of Fig. 12 was made of two parts such that 'the whole angular range from 1 mrad to about 100 mrad can be covered. The very small-angle events (in the Coulomb region) are detected by a two-arm spectrometer sharing the first four magnets with the storage ring system. The larger-angle events are momentum-analysed with a pair of magnets that do not perturb the circulating beams.

After many discussions, the ISR Committee decided to approve only the system made of two septum magnets installed in the intersection regions and to leave the detection of elastic scattering in the Coulomb region to the scintillators mounted in the Roman pots. Since then, Carlo Rubbia has described the ISR experimental program as 'key-hole physics'. After the approval, the Collaboration was joined by the Aachen and Harvard groups and became the Aachen–CERN–Harvard–Genoa–Torino (ACHGT) Collaboration.

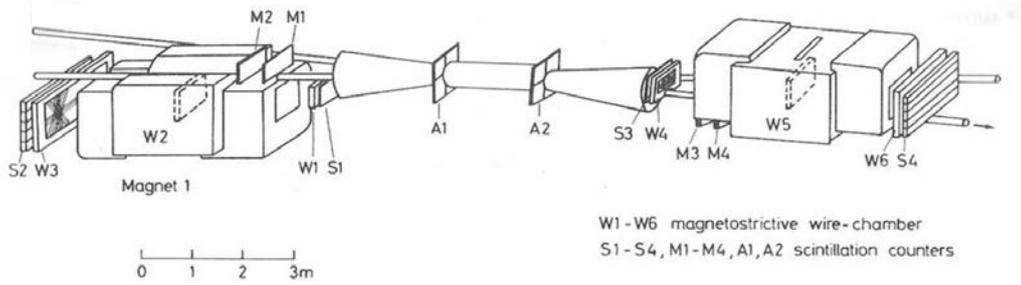

**Fig. 12:** The septum magnets of the ACHGT Collaboration, which have been used to measure the forward elastic cross-section

These three experiments were mounted in interaction regions I2 and I6 of the ISR, as shown in Fig. 13.

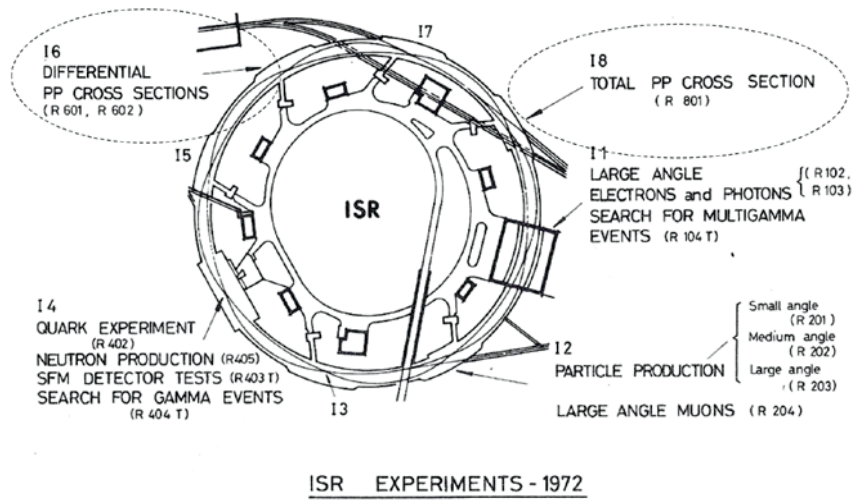

**Fig. 13:** ISR experiments in 1972: R601 = CERN–Rome, R602 = Aachen–CERN–Harvard–Genoa–Torino and R801 = Pisa–Stony Brook

A picture of the intersection region in which the ACHGT septum magnets and the Roman pots were installed is shown in Fig. 14.

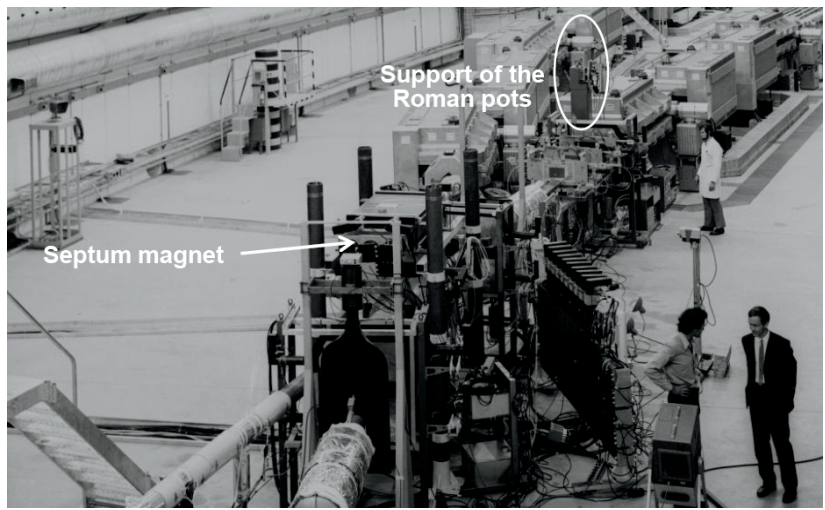

**Fig. 14:** Paolo Strolin describes to Alexander Skrinsky the ACHGT experiment, which measured with magnetostrictive spark chambers the momenta of the protons scattered between 30 and 100 mrad

# 4    First results on elastic scattering and total cross-sections

The slope of the forward elastic cross-section was the easiest measurement to perform. The 1971 results [25, 26], reported in Fig. 15, confirmed the behaviour found first at the PS and confirmed at Serpukhov: in the range $30 \leq s \leq 3000 \text{ GeV}^2$, the increment of the forward elastic slope $B$ is proportional to $\ln s$, in agreement with the description based on pomeron exchange.

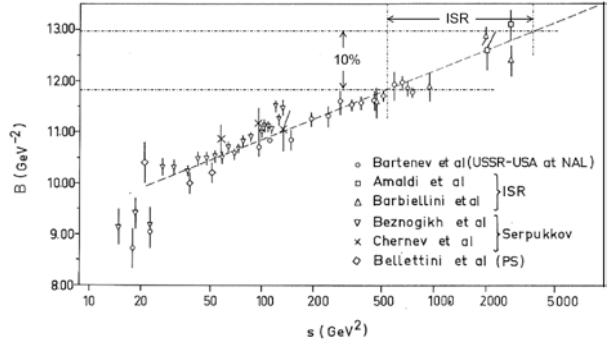

**Fig. 15:**  The data available in 1971 for $-t \leq 0.12 \text{ GeV}^2$ and the results of the measurement performed in 1972 at NAL (Fermilab) [27]. The dashed line shows that, over a very large energy range, the $t$ width (which is equal to $1/B$) of the forward elastic peak decreases as the inverse of $(a + b \ln s)$.

Taking into account all the data, the figure shows that, in the *full* ISR energy range ($23 \leq \sqrt{s} \leq 62 \text{ GeV}$, i.e., $550 \leq s \leq 3800 \text{ GeV}^2$), the slope parameter $B$ increases by about 10% on passing from $11.9 \text{ GeV}^{-2}$ to $13.0 \text{ GeV}^{-2}$, which corresponds to a 5% increase of the proton–proton interaction radius.

In the Regge description, the energy variation of $B$ is related to the slope of the pomeron trajectory at $t = 0$:

$$B = B_0 + 2\alpha'(0) \ln(s/s_0)$$

The dashed line of Fig. 15 corresponds to $\alpha'(0) = 0.28 \text{ GeV}^{-2}$, confirming what was already known from lower-energy data: the pomeron slope at $t = 0$ is definitely smaller than the slope $\alpha_\rho'(0) \approx 1 \text{ GeV}^{-2}$ of the $\rho$ trajectory shown in Fig. 2.

In 1972 the ACHGT Collaboration reported two very interesting findings [28, 29]: (i) the forward elastic cross-section has a variation of slope at $|t| \approx 0.16 \text{ GeV}^2$ (Fig. 16a) and (ii) the deep diffraction minimum located at $|t| \approx 1.4 \text{ GeV}^2$ is the energy development of the structure observed at lower energies (Fig. 16b).

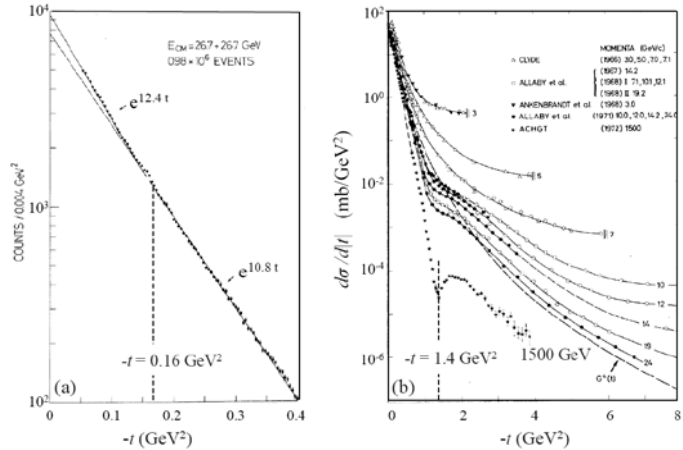

**Fig. 16:**  First measurements by the ACHGT Collaboration of proton–proton elastic scattering (a) in the forward region and (b) at large momentum transfer

However, the real surprise came with the measurements of the total cross-section done by the Pisa–Stony Brook Collaboration, with the apparatus of Fig. 6, and by the ACHGT and the CERN–Rome Collaborations, using the forward elastic cross-section and the optical theorem.

This method, which, as far as I know, was not considered before the ISR start-up, was pioneered in 1971 by ACHGT [30]: the hadron–hadron forward elastic cross-section (measured outside the Coulomb peak with the Van der Meer method) is extrapolated to zero angle to obtain $(d\sigma/dt)_0$ and the optical theorem is applied to obtain

$$\sigma_{tot} = \frac{\sqrt{16\pi\,(d\sigma/dt)_0}}{(1+\rho^2)}$$

It is worth remarking that the correction due to $\rho^2$ introduces a negligible error and that, because of the square root, the percentage error in $d\sigma/dt$ (due to the Van der Meer method and to the unavoidable errors in the measurement of the forward elastic rate) is reduced because of the square root in $\sigma_{tot}$ by a (very helpful) factor of 2.

In the autumn of 1972 the three collaborations were competing to be the first to measure the total proton–proton cross-section. I remember very vividly that period, because I was the one performing the analysis of the CERN–Rome data. In the invited talk I gave in September 1973, i.e., one year later, at the Aix en Provence International Conference on Elementary Particles [31], I summarized with a figure the confusing status of the measurements in October 1972 (Fig. 17).

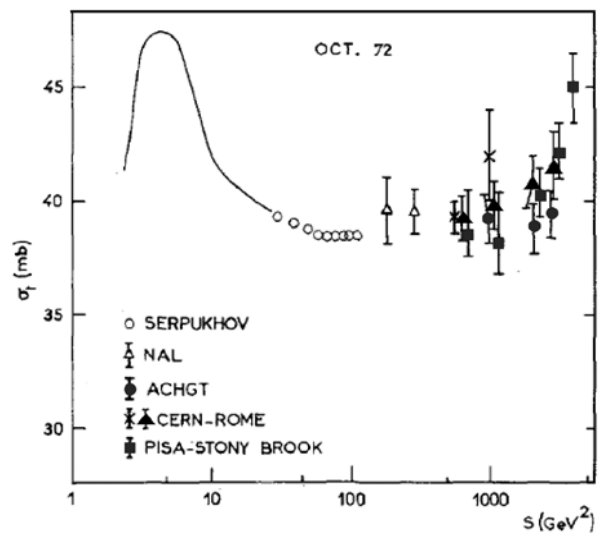

**Fig. 17:** Status of the total cross-section measurements in October 1972. The points by the CERN–Rome Collaboration were obtained with the luminosity measured with both the Van der Meer method and Coulomb scattering.

Figure 17 shows that, at that date, the Pisa–Stony Brook and CERN–Rome Collaborations had an indication of the rising cross-section, while AGHGT was finding no energy dependence. This much debated difference continued during the next months.

In February 1972 the CERN–Rome Collaboration published the first measurement of the ratio $\rho$ between the real and imaginary parts of the forward scattering amplitude and of the total cross-section using Coulomb scattering as normalization [31]. The measurement could be performed only at the two lowest ISR energies because, with the apparatus of Fig. 18a, the minimum scattering angle was fixed at about 2.5 mrad by the background rate due to the beam halo. Thus at the highest ISR energies, after completion of the stacking process in the two ISR rings, the pots could not be moved close enough to the beams to reach the $t$ range where the Coulomb scattering amplitude is as large as the nuclear one.

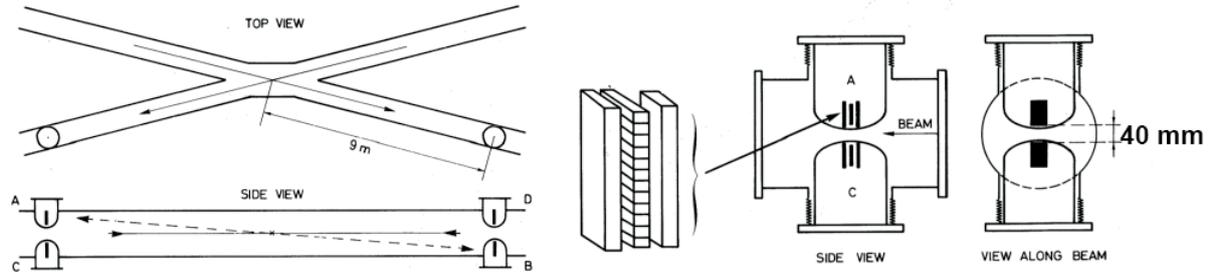

**Fig. 18:** The 1972 telescope system of the CERN–Rome Collaboration [32] was used (i) to obtain the ISR luminosity using the Coulomb scattering events and (ii) to measure ρ

The measured differential cross-sections are shown in Figs. 19a and 19b. The *t* dependence of the Coulomb amplitude is well known, because it is due to large-impact-parameter collisions of two point-like charges, is essentially real and decreases proportionally to $1/t^2$. In the *t* range indicated by the dashed ellipse, the nuclear amplitude varies little and its (small) real part interferes with the Coulomb amplitude, which is well known, being due to an electromagnetic phenomenon. The ratio *ρ* can thus be obtained by a fit to the very precise data. The results of this first experiment are shown as full dots in Fig. 19c.

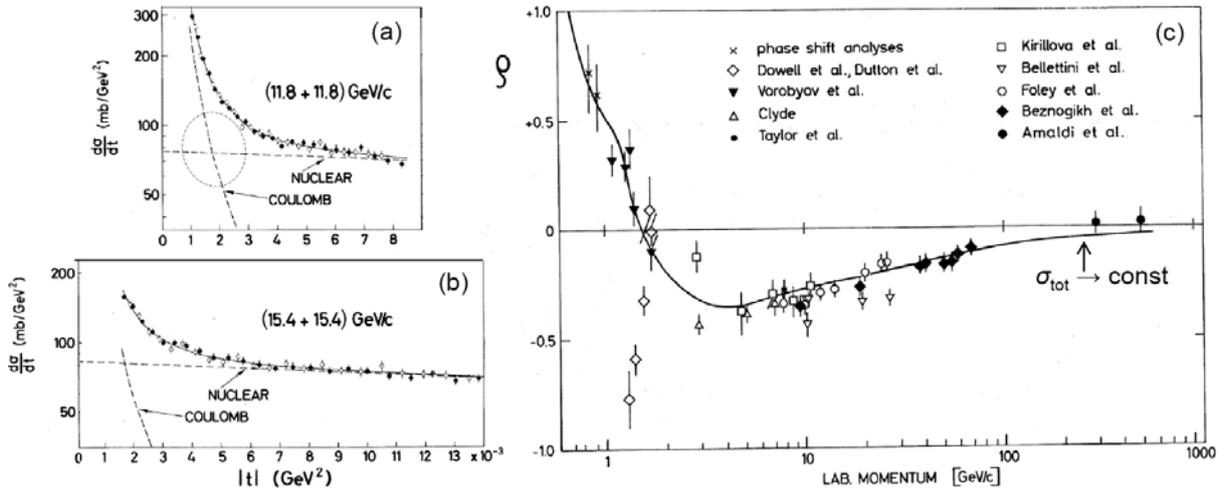

**Fig. 19:** The first measurements of the real part of the forward scattering amplitude were performed at the two lowest ISR energies [32]

The two data points indicated that *ρ* was becoming positive in the ISR energy range which, because of the Khuri–Kinoshita theorem, was a signal of the rise of the total proton–proton cross-section. The error bars are large, but within the Collaboration we knew that the indication was stronger than it appeared because, after many discussions, the errors were doubled to be on the safe side in the first paper reporting the result of a new delicate experiment.

The CERN–Rome and Pisa–Stony Brook data – presented at CERN in the already quoted March 1973 seminar and published in *Physics Letters* [33, 34] – definitely demonstrated that (i) the proton–proton total cross-section increases by about 10% in the ISR energy range (Fig. 20a) and (ii) the elastic cross-section (computed by integrating the measured differential cross-section) increases by the about same amount, so that in the full ISR energy range the ratio $\sigma_{el}/\sigma_{tot} \approx 0.17$, while it decreases monotonically at lower energies. This about constant ratio is definitely smaller than the value $\sigma_{el}/\sigma_{tot} = \frac{1}{2}$ that would result from the scattering of a wave by a black disc.

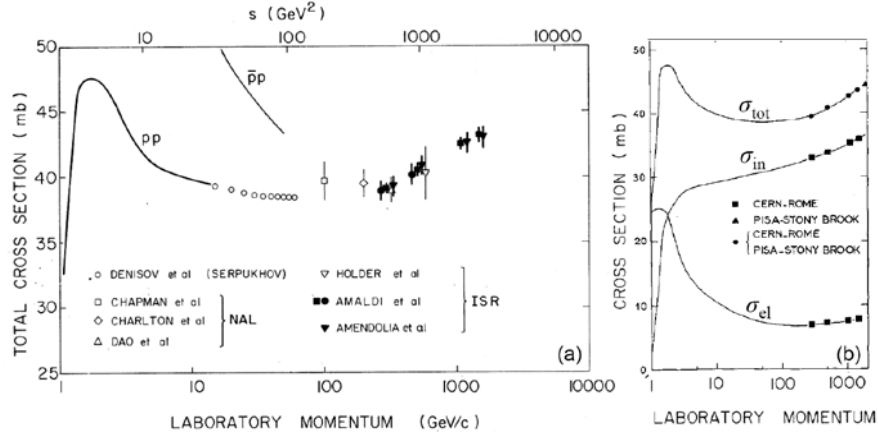

**Fig. 20:** (a) The proton–proton total cross-section increases for laboratory momenta larger than 300 GeV/*c* ($s > 500$ GeV$^2$). (b) The inelastic cross-section was computed by subtraction: $\sigma_{in} = \sigma_{tot} - \sigma_{el}$.

The inelastic cross-section is four times larger than the elastic cross-section and increases roughly proportionally to $s^{0.04}$ from about 50 MeV/*c* to the maximum ISR energy (Fig. 20b). Looking at the three curves of this figure, it appears that the shallow minimum of the total proton–proton cross-section $\sigma_{tot} = \sigma_{in} + \sigma_{el}$ around $s = 100$ GeV$^2$ is a consequence of the *continuously* rising inelastic cross-section which, through unitarity, drives the increase of the elastic cross-section.

If the energy dependence of the high-energy total cross-section is fitted with the formula of the Froissart–Martin bound, one obtains

$$\sigma_{tot} \simeq \left[ 38.4 + 0.5 \ln\left(\frac{s}{s_o}\right)^2 \right] \text{ mb},$$

where $\sqrt{s_o} = 140$ GeV [33]. Since the coefficient 0.5 mb is much smaller than the limiting value predicted by the Froissart–Martin bound, the energy dependence measured at the ISR is most probably uncorrelated with the bound itself.

As I said, at the time, most experts were convinced of the constancy of the cross-sections at high energies, with two important exceptions. In 1952 Werner Heisenberg had published a paper that described pion production in proton–proton collisions as a shock wave problem governed by a non-linear equation and deduced a $\ln^2 s$ dependence of the cross-section [35]. The model proposed by H. Cheng and T.T. Wu [36] is much more sophisticated because it is based on quantum field theory, specifically on a massive version of quantum electrodynamics. After the announcement of the ISR results, the model was reconsidered and fitted to the experimental data by Cheng, Walker and Wu [37].

The CERN seminar and, soon after, the two publications made a certain impression also outside the physics community, so much so that I wrote an article for *Scientific American*. In spring and summer 1973 this took me a lot of time since the editor was following very closely the writing of the text and the production of the figures. The article was published in September 1973 [38] after a drastic cut of the part of the article containing the impact parameter description of the ISR collision. In substitution, I introduced the quantity 'average opaqueness' $O = 2\sigma_{el}/\sigma_{tot}$, which in wave mechanics is $O = 1$ for a black disc, and showed with a figure how $O$ decreases at low energies and becomes roughly constant ($O \approx 0.35$) in the whole ISR energy range.

I also underlined that in 1972 the interest in models of rising cross-section [36] was raised when an analysis of high-energy cosmic data had indicated that the proton–proton cross-section is larger at $p \approx 10^4$ GeV/*c* than at the energies at the time available at particle accelerators [39]. However, one year later, from a different set of data, it was concluded that 'there is no evidence to suggest a change in the magnitude of the inelastic proton–proton cross-section up to 50 000 GeV' [40].

I may add that letters and telex exchanges were needed to convince the editor to insert the 29 names of the members of the CERN–Rome and Pisa–Stony Brook Collaborations, a request that in the past *Scientific American* – as they told me – had always refused because 'the readers are not interested'.

## 5   Second-generation experiments

In the years 1974–1978 three experiments brought more precise data.

The first one was performed by the Annecy–CERN–Hamburg–Heidelberg–Vienna Collaboration, which used the Split Field Magnet to accurately measure the elastic cross-section up to $q = 12$ GeV/$c$ [41]. It was observed that the minimum at $q = 1.4$ GeV/$c$ deepens around $E_{CM} = 30$ GeV and fills up at larger energies (Fig. 21a). It was interesting to remark that the deepest minimum happens at the same energies at which the forward real part is practically zero (Fig. 19c), possibly indicating that the fill-up at higher energy is due to a non-zero real part of the large-angle scattering amplitude.

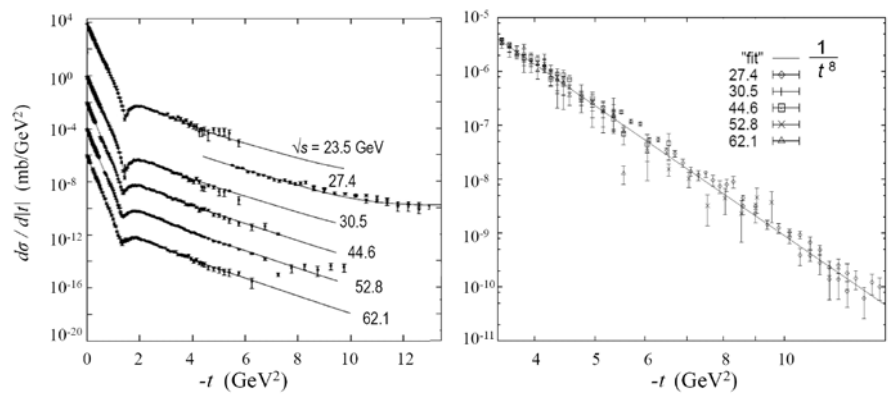

**Fig. 21:** (a) The elastic differential cross-sections at large momentum transfers plotted on different vertical scales [41]. (b) The elastic cross-section is energy independent and decreases as $1/t^8$ [42].

Figure 21b shows that the differential cross-section is energy independent when $-t$ varies in the range 3–10 GeV$^2$. The $1/t^8$ behaviour is predicted by the simple model in which the three quarks of the proton exchange a pomeron. The question [42] is this: Why does this lowest-order three-gluon exchange work so well?

Going back to 'small-angle physics', in 1973 the CERN–Rome and Pisa–Stony Brook Collaborations successfully tried a new method for measuring the total cross-section [43] and proposed to the ISR Committee a joint experiment that would have been done in new Roman pots installed – with more precise scintillator hodoscopes – in intersection region I2 where the Pisa–Stony Brook apparatus was located. Figure 22 shows the overall apparatus.

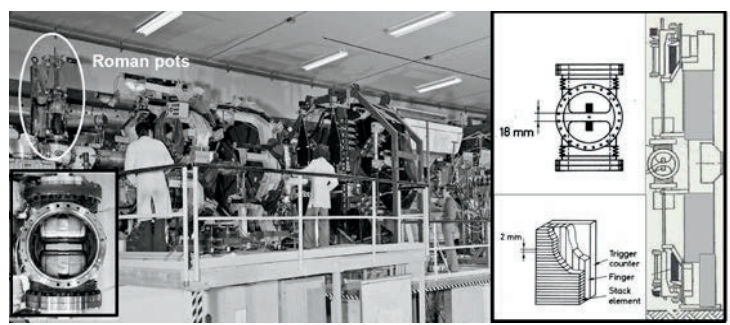

**Fig. 22:** The CERN–Rome–Pisa–Stony Brook experiment, which included two pairs of thin Roman pots (insets), was installed in I8

As the inset to Fig. 22 shows, the four pots – two per side – had very thin and flat windows, which allowed the pots – and the new systems of 'finger' scintillators they contained – to be moved much closer to the circulating proton beams, once the beam stacking process was completed. The set-up also allowed a much more accurate measurement of the distance between the edges of the two hodoscopes located one on top of the other. I well remember Giuseppe Cocconi and the NIKHEF PhD student Jheroen Dorenbosch spending long hours to improve – through accurate position measurements – the knowledge of the momentum transfer $q$.

The combination of the two detectors opened the way to the application of the new method for measuring total cross-sections. This is based on the measurement of (i) the total number of inelastic events $N_{in}$, measured by the Pisa–Stony Brook detector in a given run, which is, after small corrections due to the unavoidable losses, proportional to $\sigma_{tot}$ and (ii) the extrapolated forward rate $(dN/dt)_0$, measured by the CERN–Rome hodoscopes, which is proportional to $\sigma_{tot}^2$. Because of the optical theorem, $\sigma_{tot}$ is proportional to the ratio $(dN_{el}/dt)_0/N_{tot}$, where $(dN/dt)_0$ is the extrapolated forward number of events and $N_{tot} = N_{in} + N_{el}$ is the total number of inelastic and elastic events, computed by integrating the differential rate $dN_{el}/dt$:

$$\sigma_{tot} = \frac{16\pi}{(1+\rho^2)} \frac{(dN_{el}/dt)_0}{N_{el}+N_{in}}$$

The ratio $\rho$ is small and contributes a negligible error to the overall uncertainty.

The combined results of the three methods are plotted in Fig. 23 [44] together with the CERN–Rome measurements of the real part of the forward amplitude [45, 46] obtained with the improved Roman pots of Fig. 22. The curves have been obtained by fitting all available data and taking into account the dispersion relation, which, by neglecting spin effects, connects the forward real parts (Fig. 23b) to energy integrals of the total cross-sections (Fig. 23a).

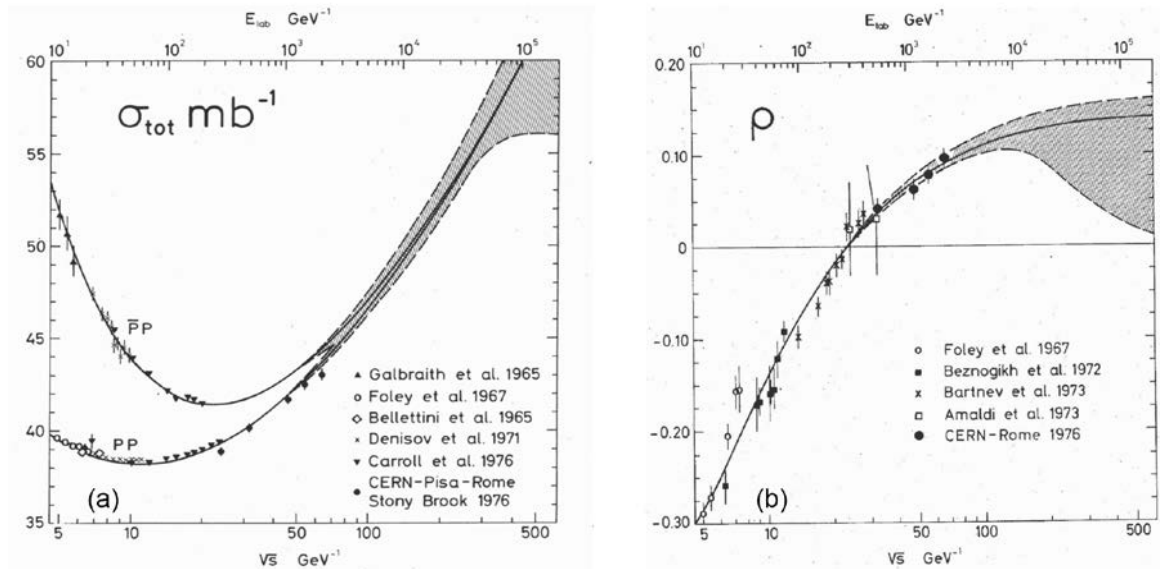

**Fig. 23:** The curves are fitted to the energy dependence of the total cross-sections and the forward real part, and are based on the analyticity properties of the scattering matrix [45, 46]

The physical content of the complicated mathematics can be understood by stating that, at high energies, $\rho$ becomes roughly proportional to the *logarithmic* derivative of the total cross-section, $d\sigma_{tot}/d(\ln s)$. This fits with the Khuri–Kinoshita theorem, which states that $\rho \to \pi \ln s$ for a cross-section that increases proportionally to $\ln^2 s$ – and explains why precise measurements of $\rho$ at $\sqrt{s} \approx 50$ GeV determine the total cross-section up to about 500 GeV. (A rigorous discussion of this very rough argument can be found in Ref. [47].) It is worth underlining that this was the first experiment in which the measured ratio $\rho$ was used to obtain information on the energy dependence of the total cross-section at energies much larger than those available in the laboratory.

The global CERN–Rome fit [45] gives a total cross-section that increases as $\ln(s/s_0)^\gamma$ with $\gamma = 2.1 \pm 0.1$ and $s_0 = 1$ GeV. The exponent coincides, within the error, with the limiting value of the Froissart–Martin bound. This fact was confirmed by a second experiment performed just before the demise of the ISR, when the availability of the CERN Antiproton Accumulator allowed a measurement of the real part of the antiproton–proton forward scattering amplitude. The CERN–Louvain-la-Neuve–Northwestern–Utrecht Collaboration used the apparatus of the CERN–Rome Collaboration and inherited most of the techniques: I remember Jheroen Dorenbosch and myself passing to Martin Bloch the codes we had developed over the years.

The experiment was a success and confirmed the proton–proton results reported in Fig. 23b [48]. More importantly, the antiproton–proton forward real part was measured to be positive, albeit with larger errors, as expected – due to the Pomeranchuk theorem – for a rising cross-section that becomes asymptotically equal to the proton–proton one. An overall fit with $s_0 = 1$ GeV, which included preliminary data obtained at the CERN proton–antiproton collider, gave $\gamma = 2.02 \pm 0.01$.

To understand the significance of these results, let us go a step backwards.

By applying to the scattering amplitude $f(t)$, which is a function of $q = (-t)^{1/2}$, the transformation written in Fig. 24, one can compute the 'profile function' $\Gamma(a)$ as a function of the impact parameter $a$ in the plane perpendicular to the momenta of the colliding particles.

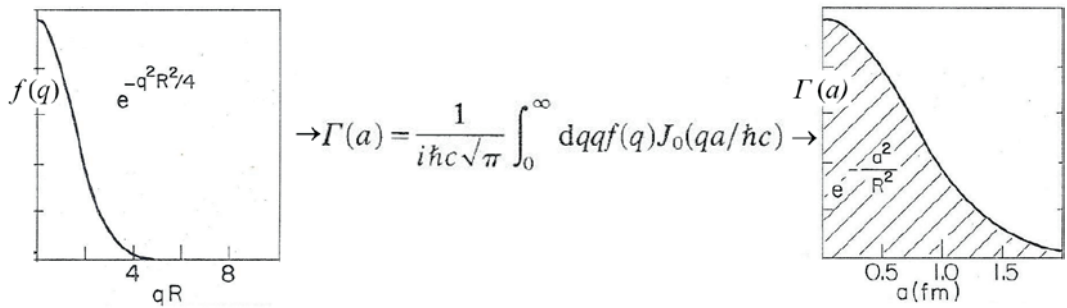

**Fig. 24:** A Gaussian elastic profile function corresponds to a scattering amplitude that decreases exponentially with $q^2 = |t|$. (In the integral, $J_0$ is the Bessel function of order zero.)

By applying the same transformation to the three terms of the unitarity relation, one obtains

$$2\,\mathrm{Re}\,\Gamma(a) = |\Gamma(a)|^2 + G_{\mathrm{in}}(a) \quad \text{with} \quad 0 \le \Gamma(a) \le 1 \quad 0 \le G_{\mathrm{in}}(a) \le 1$$

where the two inequalities express the limits imposed by unitarity. This equation shows how, in the diffraction limit, i.e., when $f(q)$ is essentially imaginary because $\rho$ is small and $\Gamma(a)$ is practically real, the *inelastic overlap integral* $G_{\mathrm{in}}(a)$ determines the elastic profile function, $\Gamma(a) = 1 - \sqrt{[1 - G_{\mathrm{in}}(a)]}$, and vice versa, so that $\Gamma(a)$ and $G_{\mathrm{in}}(a)$ can be obtained by applying the Bessel transformation to the measured scattering amplitude $f(q)$.

If the inelastic overlap integral equals 1 up to an impact parameter $a = R$, the same happens to the profile function, which thus describes a black disc. In this case the elastic and inelastic cross-sections, given by the integrals of $|\Gamma(a)|^2$ and $G_{\mathrm{in}}(a)$, are equal ($\sigma_{\mathrm{el}} = \sigma_{\mathrm{in}}$) so that $\sigma_{\mathrm{tot}} = \sigma_{\mathrm{el}} + \sigma_{\mathrm{in}} = 2\sigma_{\mathrm{el}}$ and $\sigma_{\mathrm{el}}/\sigma_{\mathrm{tot}} = 0.5$, as mentioned above. In the ISR energy range, this ratio is $\sigma_{\mathrm{el}}/\sigma_{\mathrm{tot}} = 0.17$ and the colliding protons are not black but transparent to one another.

This statement can be made quantitative by computing $\Gamma(a)$ and $G_{\mathrm{in}}(a)$ from the measured elastic differential elastic cross-sections [49]. Figure 25a shows that the profile function is Gaussian-like and completely different from that of Fig. 25b, which describes a black disc having a radius proportional to $\ln(s/s_0)$ and a grey periphery of constant width, as needed to saturate the Froissart–Martin bound. (It can be noted that this is the high-energy behaviour predicted by the Cheng and Wu massive quantum electrodynamics model [36, 37].)

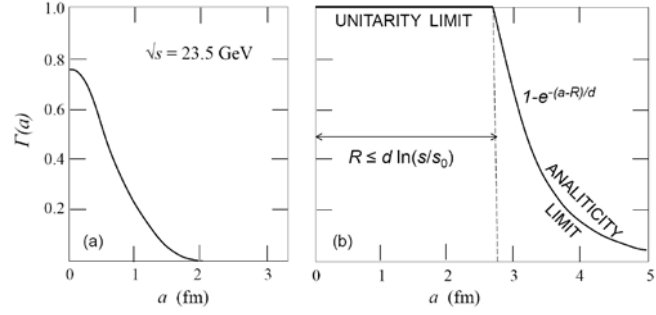

**Fig. 25:** At ISR energies the profile function [49] is far from saturating the unitarity and analyticity constraints that define the Froissart–Martin bound

Figure 25b is taken from Ref. [43], where the impact parameter descriptions of the analyticity limit and of the asymptotic behaviour of the forward real part are discussed in detail. In the simplified version depicted in Fig. 25b, the Froissart–Martin limiting profile function contains the length $d$, which is determined by the pion mass and fixes the maximum constant $C$ that asymptotically multiplies $\ln^2(s/s_0)$. In the original works of the 1960s [6, 7], $C$ was proven to be equal to $\pi(\hbar/m_\pi)^2 \approx 60$ mb, but in a 2009 paper [50] André Martin derived the new limit $C = \pi(\hbar/2m_\pi)^2$, which is four times smaller and corresponds to $d = \hbar/[(2\sqrt{2})m_\pi] \approx 0.5$ fm. It is worth noting that the new constant $C \approx 15$ mb is still thirty times larger than the best fit to the experimental data.

I now consider the measured *increase* $\Delta G_{in}(a)$ of the inelastic overlap integral over the ISR energy range. In 1973 I presented such an analysis in Aix en Provence, concluding that the increase of the proton–proton cross-section is a *peripheral* phenomenon [31], a conclusion reached at the same time by others [5051a, 51b].

This is confirmed by Fig. 26a, which is the result of an analysis performed in 1980 with Klaus Schubert on *all the data* collected at the ISR [49]. The novelties brought by this analysis were the direct calculation of $G_{in}(a)$ from the experimental data and a careful estimate of the effects of statistical and systematic errors. Figure 26b displays the results of the analysis by Henzi and Valin [52], who used a different approach by first fitting the differential cross-sections with analytical functions and then computing $G_{in}(a)$.

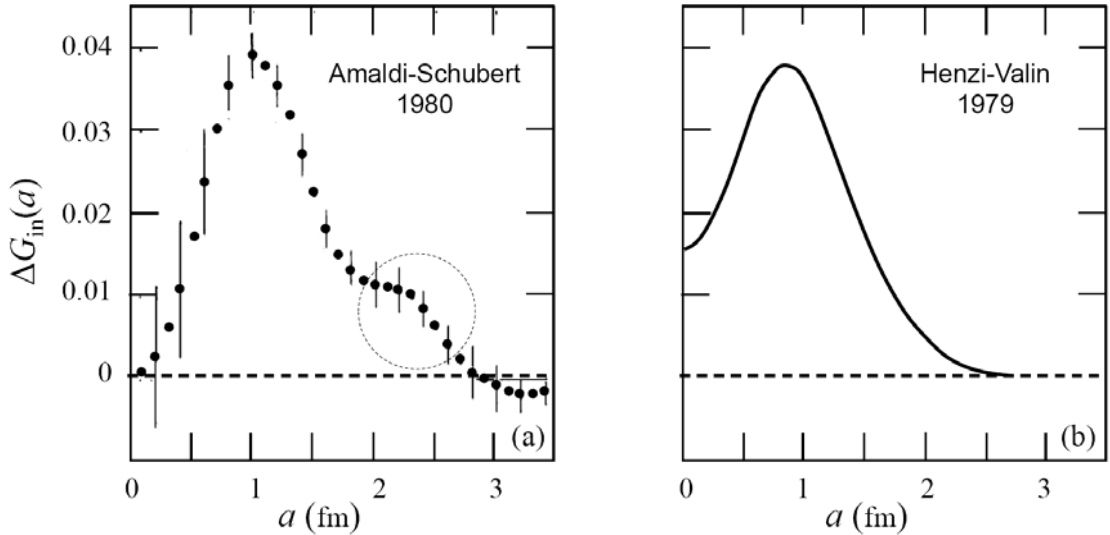

**Fig. 26:** The variation of the inelastic overlap integral in the ISR energy range (23 GeV $\leq \sqrt{s} \leq$ 62 Gev) as a function of the impact parameter $a$ (expressed in fermis) is a good way to describe the physical significance of the phenomenon of the rising total proton–proton cross-section with energy

It is seen that the shadow of the inelastic channels increases by $\Delta G_{in} = 0.04$ at 1 fm, which confirms the peripheral nature of the phenomenon. At $a = 0$ the two analyses are compatible when the errors are properly taken into account and indicate that $\Delta G_{in}(0)$ is less than three times smaller than $\Delta G_{in}(1\text{ fm})$. It could even be zero, since small impact parameters imply large moment transfers, and in this region the analytical fits to the cross-section [52] are not perfect, a problem that is not encountered when the experimental data are used directly [49]. It is also worth mentioning that the physical origin of the bump of $\Delta G_{in}(a)$ in Fig. 26a at $a = 2.3$ fm is not known.

As mentioned above, the fitted exponent of the logarithmic increase of $\sigma_{tot}$ is 2, with a very small error. We can now answer the question: Is this fact connected with the exponent 2 predicted by the Froissart–Martin bound? The answer must be negative, because the overlap integral of Fig. 25a is very different from that of Fig. 25b, but the coincidence is so puzzling that, without understanding, the expression 'qualitative saturation of the Froissart–Martin bound' was introduced and much used.

As the last argument of this section, let us consider the $t$-channel description of diffractive scattering from the impact parameter point of view. By applying the Bessel transformation to the pomeron amplitude with $\alpha(0) = 1$, one obtains a profile function $\Gamma(a)$ that has a radius $R$ that *increases* with energy as $\ln(s/s_0)$ and a central value that decreases. Thus in the Regge model with pomeron intercept $\alpha(0) = 1$, the forward peak *shrinks* as $\ln(s/s_0)$ while the central value *decreases* as $\ln(s/s_0)$, so that the total cross-section remains *constant*. This is certainly not in agreement with the measurements summarized by the function $\Delta G_{in}(a)$ represented in Fig. 26.

In synthesis, the 1973 ISR measurements of elastic scattering and total cross-section highlighted an unexpected state of affairs: with increasing collision energy, the proton–proton 'opacity' at zero impact parameter does not decrease – as predicted by the 'classical' pomeron exchange model – but remains about constant.

## 6    Particle production and diffraction dissociation

The first experiments performed at the ISR on particle production observed the two main properties well known at lower energies: the transverse momenta were small and about half of the total energy $\sqrt{s}$ was going, on average, in the forward direction, giving rise to what was called the 'leading particle effect'.

The two variables used to describe the inclusive production of single particles are the fractional momentum $x = 2p_L/\sqrt{s}$ (where $p_L$ is the longitudinal momentum in the centre-of-mass system) and the rapidity $y = \tfrac{1}{2}\ln[(E + p_L)/(E - p_L)]$, where $E$ is the total energy of the particle. The importance of the variable 'rapidity' stems from the fact that, in non-relativistic kinematics, the rapidity of a particle coincides with its velocity and, in the relativistic regime, rapidities add linearly – as do non-relativistic velocities – while velocities do not.

When the production angle of a relativistic particle in the centre-of-mass system is $\theta = 0$, so that $p_L = p = \sqrt{(E^2 - m^2)}$, the two variables vary within the ranges

$$-1 \leq x \leq 1 \quad \text{and} \quad -y_{max} \leq y \leq y_{max} \quad \text{with} \quad y_{max} = \tfrac{1}{2}\ln(s/m^2)$$

Thus the maximum rapidity in a proton–proton collision is $\ln[(\sqrt{s})/m_p]$: it was 2 at the PS ($\sqrt{s} = 6.8$ GeV) and became 4.2 at the maximum ISR energy ($\sqrt{s} = 63$ GeV).

At the Aix en Provence Conference of September 1973, Giorgio Bellettini and Lorenzo Foà presented the most recent data obtained by the Pisa–Stony Brook Collaboration by showing, among other results, the plots reproduced here as Fig. 27a [53]. Since the hodoscopes of Fig. 8 measured the angles $\theta$ of the outgoing particles with respect to the beam direction, and not the energy, the pseudorapidity $\eta = -\ln\tan\theta/2$ was used. (Note that $y = \eta$ for massless particles and that, for particle with mass, $y \rightarrow \eta$ at high energies and for most angles, but not in the very forward region, since for $\theta \rightarrow 0$ the pseudorapidity increases without limit.)

After subtracting the two tracks having maximum and minimum pseudorapidity, each recorded event was plotted (for a given multiplicity $n_{ch}$ of the observed charged tracks) as a point having coordinates $\eta_{av}$ and $\delta(\eta_{av})$, which are the *average* pseudorapidity of the remaining tracks and the *dispersion around* the average. The surfaces were drawn as smooth interpolations of the data.

The events clearly subdivide in two classes: (i) the 'central' inelastic events, which dominate for large multiplicities and cluster around $\eta_{av} = 0$ with large dispersion $\delta(\eta_{av})$, and (ii) the 'forward' inelastic events, which dominate at low multiplicities, have an average pseudorapidity close to the maximum (and the minimum) and a small dispersion.

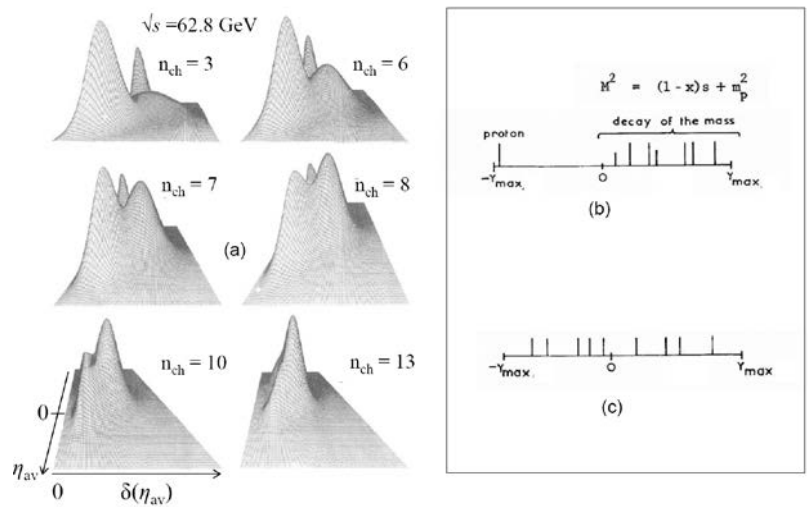

**Fig. 27:** (a) Distribution of the events measured by the Pisa–Stony Brook Collaboration at the maximum ISR energy as a function of the charge multiplicity $n_{ch}$ [51]. The coordinates represent the average pseudorapidity and the dispersion around the average. (b,c) Rapidity distributions of the particles (b) in a diffraction dissociation event and (c) in a central inelastic event.

The second class is dominated by *single diffraction dissociation* events of the type represented along the y-axis of Fig. 27b: on one side there is the proton, which has a large fractional momentum $x$, and on the opposite side there are a few particles, which have an invariant mass $M$ such that $M^2 = (1 - x)s + m^2$, where $m$ is the proton mass.

At PS energies the phenomenon of single diffraction dissociation with production of the first excited states of the proton had been well measured [54]. The new features, discovered at the ISR by the CERN–Holland–Lancaster–Manchester (CHLM) Collaboration with the detector shown in Fig. 28, was a highlight of ISR small-angle physics.

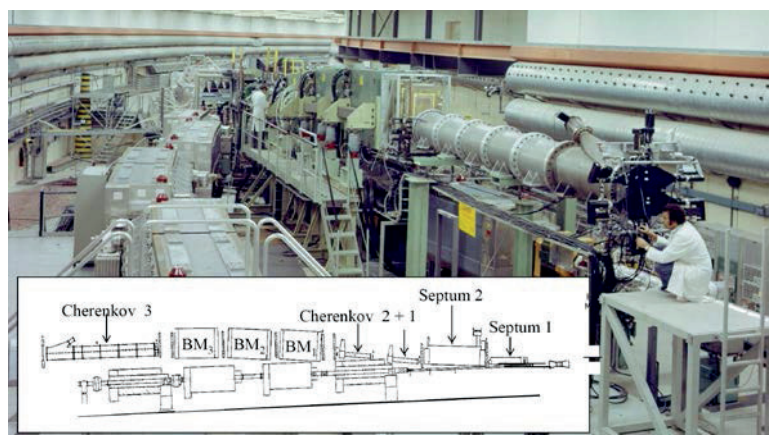

**Fig. 28:** As shown in Fig. 13, the apparatus of experiment R201 by the CERN–Holland–Lancaster–Manchester Collaboration was mounted in intersection region I2

By accurately measuring the momentum $p$ of the forward-going proton, the mass $M$ of the system moving in the opposite hemisphere and the momentum transfer $t$ could be computed. As early as 1973 the CHLM Collaboration concluded [55] that the *invariant* cross-section shows a peak for $x \approx 1$ that *does not* change when the collision energy increases (Fig. 29a). (Note that most of the events belonging to the quasi-elastic peak ($0.95 \leq x \leq 1$) correspond to excitation of states with large masses: up to $M = 10$ GeV for the top ISR energy.) Figures 29b and 29c give other important properties of large-mass diffraction dissociation discovered by the CHLM Collaboration in the following two years [56, 57].

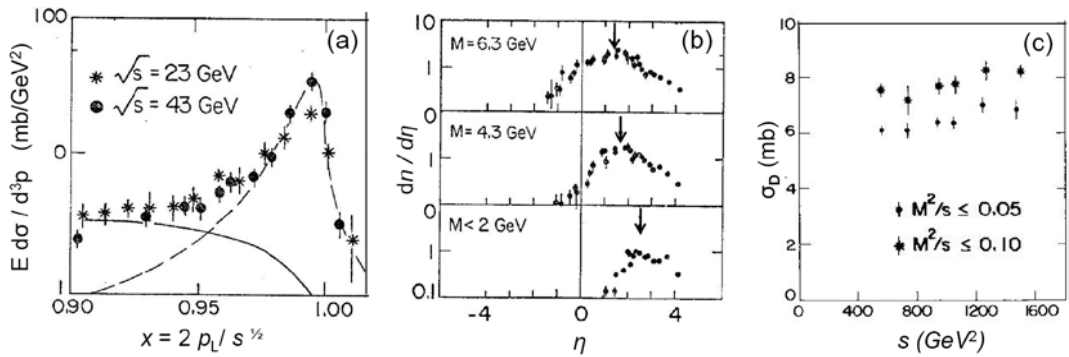

**Fig. 29:** (a) The high-mass peak of the invariant single diffraction cross-section, for a transverse momentum equal to 0.525 GeV/$c$, is energy independent [55]. (b) For all masses the pseudorapidity distributions peak around the arrows, which indicate the centre of the distribution expected from kinematics [56]. (c) The integrated single diffraction dissociation cross-section $\sigma_D$ has a slight energy dependence in the range 550 GeV$^2 \leq s \leq$ 1500 GeV$^2$ [57].

In the $t$-channel approach, this phenomenon is interpreted by drawing a pomeron exchange graph (Fig. 30a′) similar to the one describing diffractive elastic scattering. This justifies the name 'single diffraction dissociation' of one of the incoming protons into a system of mass $M$. The rapidity span is $\ln(s/mM)$ (Fig. 30a) and $M$ must have the same quantum numbers as the incoming proton, while spin and parity may be different because orbital angular momentum can be transferred by the exchanged pomeron. The three phenomena depicted in the figure are characterized by large 'rapidity gaps' and are particularly interesting when systems of large mass ($M > 2.5$ GeV) are produced.

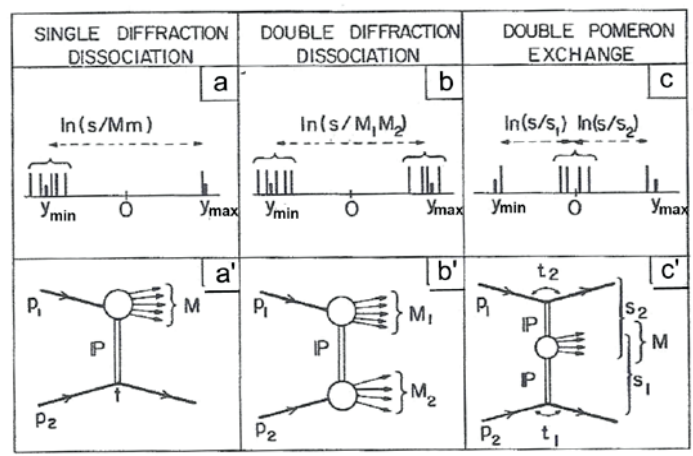

**Fig. 30:** Typical rapidity configurations of (a,a′) single diffraction dissociation, (b,b′) double diffraction dissociation and (c,c′) double pomeron exchange [58]

In the following years, the phenomenon was further studied, and in 1976 the CHLM Collaboration published other results, the most important being the measurements of the single diffraction cross-section $\sigma_D$ [57], reproduced in Fig. 29c, which shows that its value (7–8 mb) is similar to that of the elastic cross-section.

The data were not sufficiently precise to decide whether the single diffraction cross-section also increases in the ISR energy range. However, the experimental fact [59] that (at fixed $s$ and $t$) the invariant cross-section $d\sigma_D/dM^2$ *decreases* as $1/M^2$ indicates that its integral *increases* as $\ln s$, since it has to be computed for $0.95 \leq x \leq 1$, i.e., up to a maximum value of $M^2 = 0.05s$, which increases linearly with $s$. This and other interesting aspects of diffraction dissociation are discussed in two review papers published in 1976 and 1981 [58, 60].

Experimentally, *double* diffraction dissociation (Fig. 30b′) is much more difficult to study than single diffraction because the reconstruction of the two masses $M_1$ and $M_2$ requires both a large enough total rapidity span and the measurement of charged and neutral particles in the two forward cones. These conditions were not quite satisfied at the ISR.

Still, an estimate of the double diffractive cross-section could be made using ISR and Fermilab data, so much so that K. Goulianos, in the section 'Elastic and total cross-sections – Are they related through diffraction dissociation?' of a very often quoted review paper published in the closing year of the ISR [61], argued that the 'peripheral' rise of the total cross-section discovered at the ISR could be driven exclusively by the rapid increase of the sum of single and double diffraction dissociation.

The focus of ISR small-angle physics on the pomeron brought to light the very important phenomenon of Fig. 30c′, the so-called 'double pomeron exchange', which deserves a short discussion even if it is not really 'small-angle physics'.

In this reaction – even to produce a low-mass state having the quantum numbers of the vacuum – the two rapidity gaps have to be larger than $\Delta y = 3$ and the two final protons must have $x > 0.95$, conditions that are satisfied only at the largest ISR energies. In the last days of the ISR, forward drift chambers were added to the Axial Field Spectrometer (AFS) and the AFS Collaboration collected high-statistics data on the production of pion pairs in pomeron–pomeron collisions, as well as observing two kaons, four pions and proton–antiproton central states [62].

In a recent review paper on double pomeron exchange, Albrow, Coughlin and Forshaw [63] discussed the evolution of the field from the ISR times to the expectation that a single Higgs particle could be produced in Large Hadron Collider (LHC) pomeron–pomeron collisions. Such a discovery would be the last pillar of a long bridge whose first pillar was built forty years ago on the ISR shore.

Multiparticle production is the final subject to be discussed with reference to the very interesting results of the long-standing activity of the CERN–Bologna–Frascati (CBF) Collaboration working at the Split Field Magnet. The presentation will be short because, on the occasion of the ISR fortieth anniversary, the subject has been well covered by Antonino Zichichi [64, 65]. More details can be found in Ref. [66].

As already mentioned, at the ISR the bulk of the particles created in the collisions have small transverse momenta (of the order of 200 MeV/$c$, which correspond to a source having a radius of about 1 fm) and also a uniform rapidity distribution. The mean multiplicity of the events was expected to increase almost logarithmically with energy, a feature that was duly confirmed in the first months of running.

At the time, this mean multiplicity, when plotted as a function of the centre-of-mass energy $\sqrt{s}$, was different from that measured in electron–positron collisions, and nobody had thought to look for a correlation. Years later, the detailed study of thousands of such events guided the CBF Collaboration to the definition of an 'effective energy', which takes into account the fact that in the ISR events a large fraction of the energy of the colliding protons is taken away by the leading baryons, as shown in Fig. 31a. The situation is very different in electron–positron annihilations (Fig. 31b) and in deep inelastic scattering (DIS) induced by either charged leptons or neutrinos (Fig. 31c).

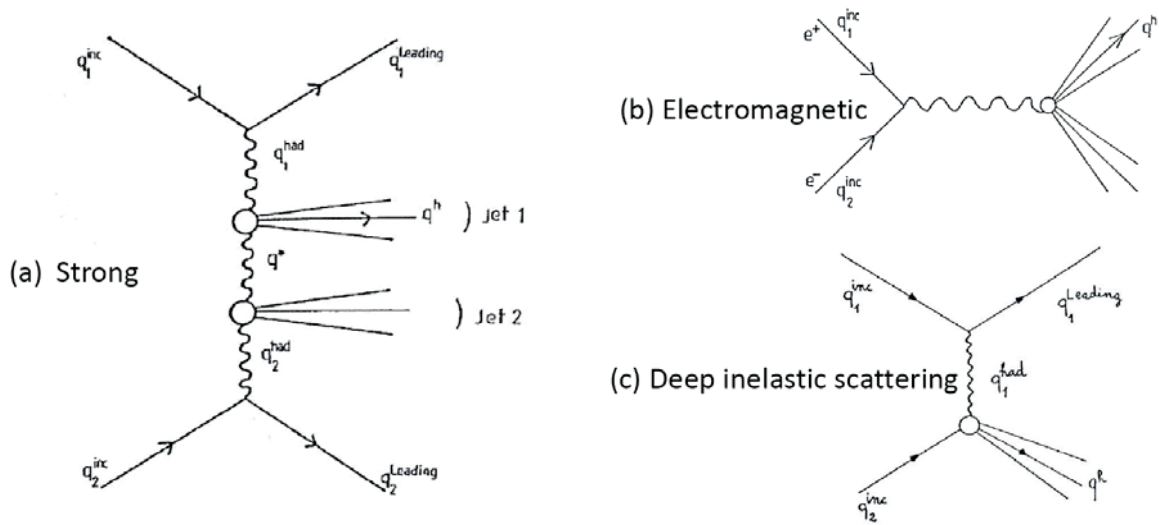

**Fig. 31:** In these kinematical graphs the quantities $q$ are four-vectors. The final hadron jets are due to (a) the strong interaction, (b) the electromagnetic interaction in the time-like region and (c) either the electromagnetic or the weak interaction in the space-like region.

On subtracting the leading particles, the effective energy in the centre-of-mass system (indicated by the authors with the symbol $2E_{had}$) is determined by the four-vectors $q_{1,2}^{had}$, so taking into account the leading particle effect. In the same reference system, one can also compute the $x$ variable of a hadron which (as shown in Fig. 31a) has momentum $q^h$. As shown in Fig. 32, a continuum of effective energies contribute to the distributions measured in collisions that happen at a given 'nominal' proton–proton energy (for instance $\sqrt{s} = 62$ GeV).

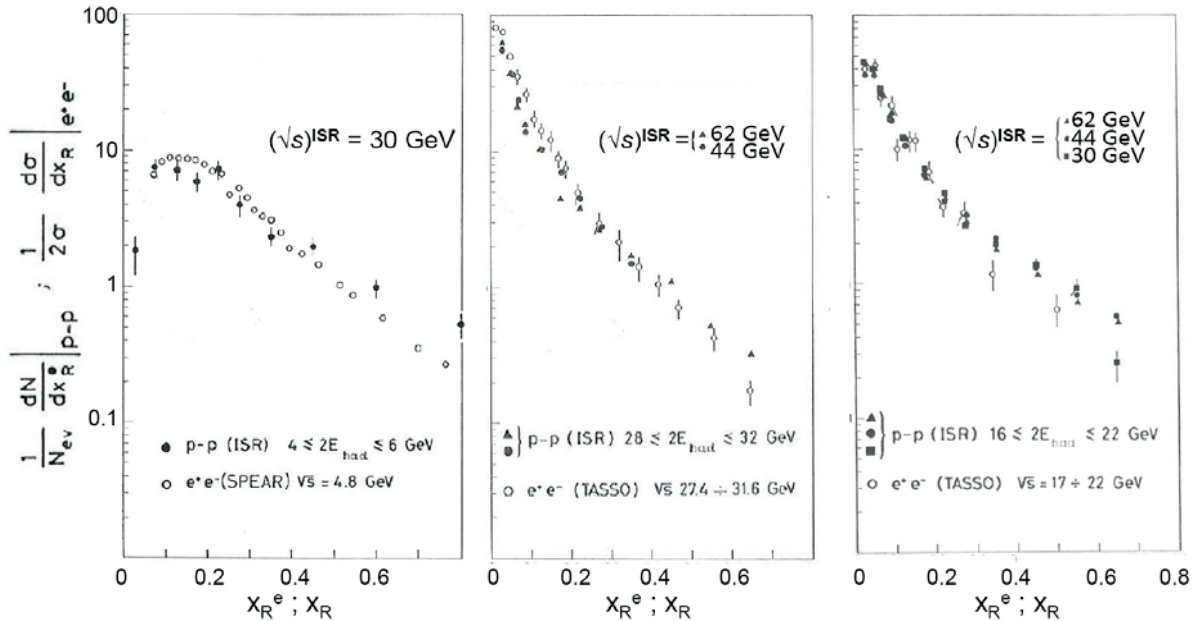

**Fig. 32:** The $x$ distribution of the hadrons produced in pp collisions and in electron–positron annihilation are practically identical when the proper effective energy is used to classify the events

In Fig. 32 the $x$ distributions of the single hadrons produced at the ISR – subdivided into three energy bands of effective energy – are compared with the results obtained in electron–positron collisions at the corresponding total energies.

Since the single-particle *x* distributions are similar, as shown in Fig. 32, the mean multiplicities in proton–proton and electron–positron collisions have the same energy dependence when the effective energy is used as independent variable.

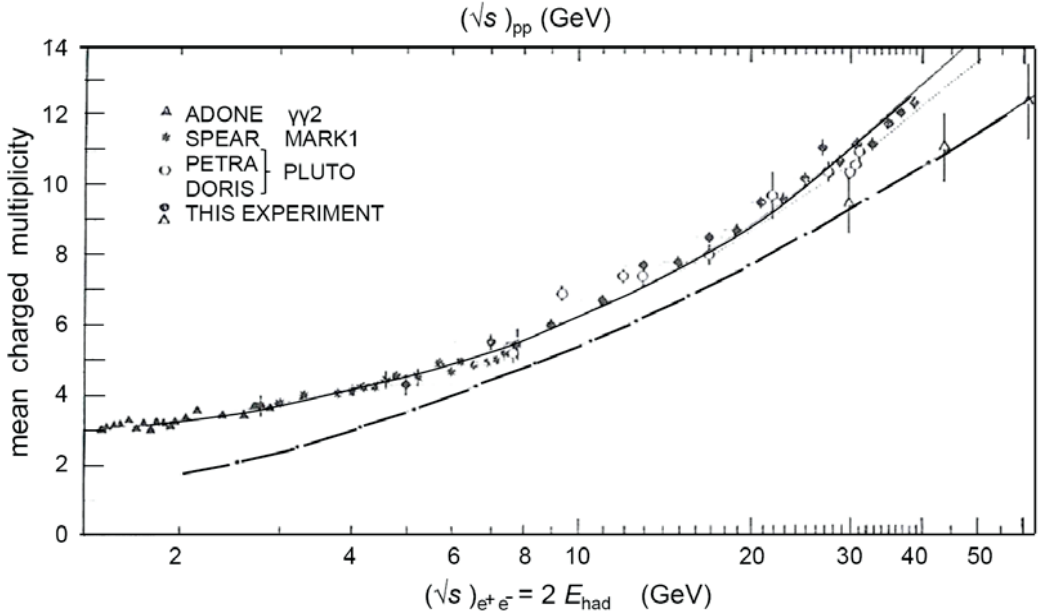

**Fig. 33:** As discovered by the CERN–Bologna–Frascati Collaboration, the mean multiplicities of electron–positron and proton–proton events cluster around the same continuous line when plotted versus the effective energy. The dash-dotted line represents proton–proton mean multiplicity plotted versus the collision energy $\sqrt{s}$.

Similar 'universality features' have been found by the CBF Collaboration when comparing, as a function of the effective energy, experimental data concerning the average charged-particle distributions in pp and νp collisions, the transverse momentum distributions and even scale breaking effects in pp and $e^+e^-$ collisions. Moreover, when a hadron is present in the initial state, the leading effect is also universal and is determined by the 'flow' of quarks from the initial to the final state.

## 7   The ISR 'small-angle physics' seen from higher energies

In forty years, the energy of hadron–hadron colliders has passed from $\sqrt{s} = 30$ GeV, the ISR minimum value, to the $\sqrt{s} = 7000$ GeV available at the LHC starting in 2009. A review of the first results from this high-energy frontier is beyond the scope of the present paper, which however closes with some remarks concerning the energy evolution of the main phenomena discussed in the previous sections.

As a first point, let me consider multiple particle production and the physical quantities discussed at the end of the last section: it will be most interesting to compare the distributions of the many quantities studied at the ISR with the data collected at the LHC (pp collisions) and the Large Electron–Positron (LEP) collider (electron–positron annihilation). A confirmation of the universality features observed at energies that are ten times larger would give even more weight to the concepts of 'effective energy' and of 'quantum number flow'.

Going back to total cross-sections and real parts of the forward scattering amplitude, Fig. 34 reproduces the data obtained at the CERN antiproton–proton collider and at the Tevatron. The LHC total cross-section measurement published in 2011 by the TOTEM Collaboration [67] has been added to the summary figures, which describe all the data up to the Tevatron energy [68]. The best fit passes through the LHC point and gives $\gamma = 2.2 \pm 0.3$ as the exponent of the $\ln s$ term, in good agreement with what was found at the ISR [45].

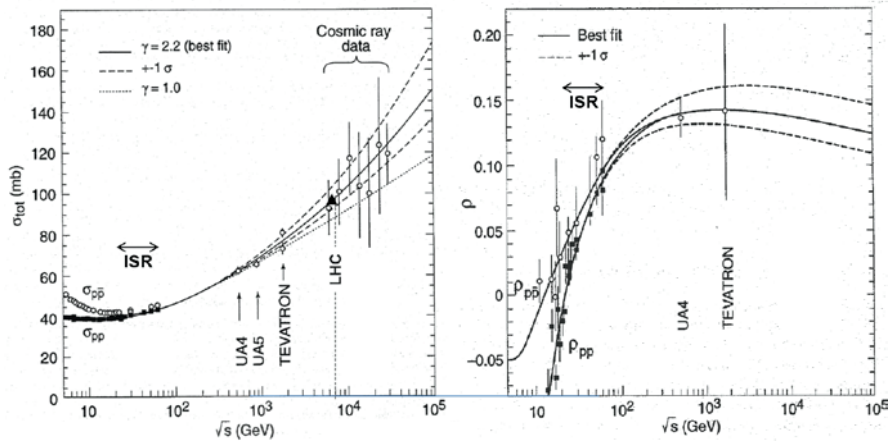

**Fig. 34:** The continuous lines represent the best fits to all the data, excluding the total cross-section measured in 2011 at the LHC: (98.3 ± 0.2 ± 2.8) mb [65]

Also, the slope of the forward elastic cross-section continues the trend measured at lower energies (Fig. 35), so that, by interpreting it as due to the slope of the pomeron trajectory, one still obtains $\alpha'(0) = 0.25$ GeV$^{-2}$.

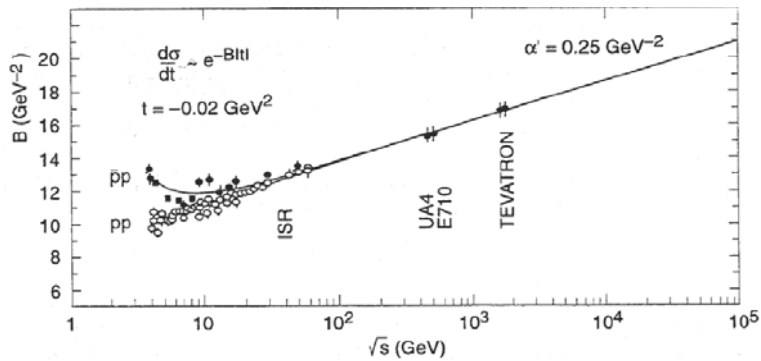

**Fig. 35:** The forward elastic cross-section shrinks as $\ln s$ in an enormous energy range: $30 \leq \sqrt{s} \leq 2000$ GeV

As far as the elastic and single diffraction dissociation cross-sections are concerned, Fig. 36 (taken from the review paper by Giorgio Matthiae [69]) shows that the single diffraction cross-section rises with energy in the same energy range (30–2000 GeV) and that the ratios $\sigma_{el}/\sigma_{tot}$ and $\sigma_D/\sigma_{tot}$, which are equal and constant in the ISR energy range, diverge at larger energies.

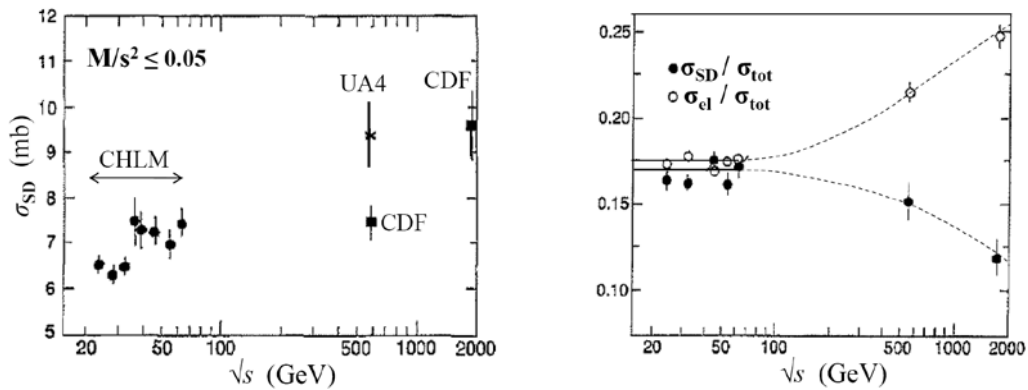

**Fig. 36:** The energy dependence of the cross-section for single diffraction dissociation $\sigma_D$ seems to increase with energy (a) but the ratio $\sigma_D/\sigma_{tot}$ definitely decreases (b), so that the importance of the phenomenon reduces at high energies

Figure 35b clearly indicates that the ISR energy range, in which masses larger than about 2.5 GeV can be produced in single diffractive dissociation, is a *transition region* and that the constancy of the ratio $\sigma_{el}/\sigma_{tot}$ with energy is not an asymptotic behaviour.

Nevertheless, the scaling with $1/M^2$ of the single diffractive cross-section, found at the ISR, still holds at 500 GeV, as shown in Fig. 37, in agreement with the prediction of the triple pomeron exchange model.

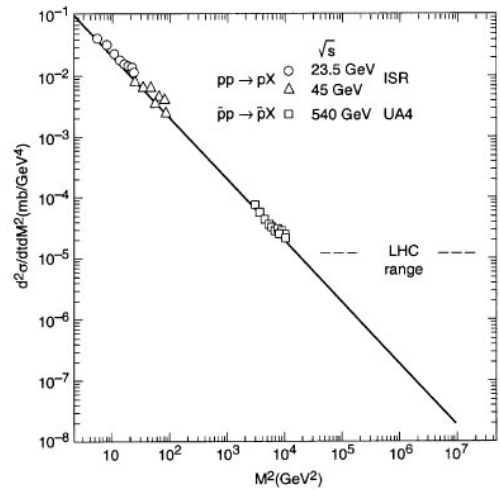

**Fig. 37:** The invariant single diffraction cross-section at a fixed $t$ value ($-t = 0.5$ GeV$^{-2}$) measured at the CERN antiproton collider scales as $1/M^2$ as found at the ISR [69]

It is clear that the LHC, with its very large centre-of-mass energy, opens new possibilities because systems with very large masses can be excited by single (and also double) diffraction dissociation; for $M^2 \leq 0.05s$, the mass $M$ of a singly diffracted system can be as large as 1.6 TeV!

Let us now consider elastic and total cross-sections. In the ISR energy range, the almost constant ratio $\sigma_{el}/\sigma_{tot} \approx 0.17$ (Fig. 36b) was an indication of an energy-independent value of the central inelastic overlap function $G_{in}(a = 0)$, even if this is not a rigorous conclusion, as shown by the analysis reported in Fig. 26b. Since the ratio $\sigma_{el}/\sigma_{tot}$ increases at higher energies, as shown in the same figure, it does not come as a surprise that the behaviour of $\Delta G_{in}(a)$, in passing from the ISR to the CERN proton–antiproton collider, definitely increases with energy as shown by Henzi and Valin [70].

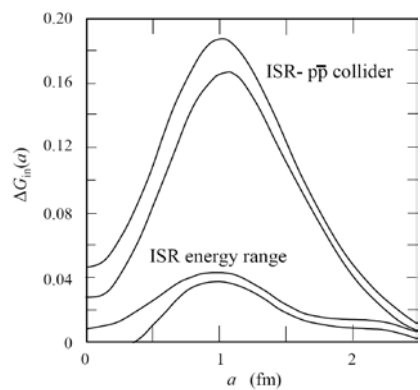

**Fig. 38:** (a) In the energy range that goes from the ISR to the CERN proton–antiproton collider (i.e., from 53 to 550 GeV), the central inelastic overlap integral increases [70]. (b) In the ISR energy range (i.e., from 23 to 62 GeV), the errors are such that no definite conclusion can be drawn on $G_{in}(0)$ [49]. The bands represent the estimated errors.

In summary, the *s*-channel description based on a *purely peripheral* increase of the inelastic overlap integral (sometimes called 'geometrical scaling') may be valid, but it is not certain, in the ISR energy range. This, together with the decreasing importance of single diffraction dissociation (Fig. 36b), implies that the rise with energy of the total cross-section may be driven by single and double diffraction up to √s ≈ 100 GeV, but this is not the case at higher energies.

I conclude this discussion of the *s*-channel description of high-energy scattering by recalling that Henzi and Valin gave to their 1983 paper [70] a well-chosen title: 'Towards a blacker, edgier and larger proton'.

As a final argument, I consider the complementary, *t*-channel description of the energy dependence of hadron–hadron cross-sections.

In 1992 Donnachie and Landshoff wrote all the hadron–hadron total cross-sections as the sum $\sigma_{tot} = Xs^\varepsilon + Ys^{-\eta}$ of two powers, the first being due to pomeron exchange and the second to the exchange of the trajectory of Fig. 2 [3]. Figure 39 shows the experimental points and the fitted curves for the four best-measured channels. Of course, all known channels were included in the fit and the model contains 15 free parameters, most of which describe the low-energy behaviour of the cross-sections. It has to be stressed that the intercept of the reggeon trajectory ($\alpha_R(0) = 0.45$) is in good agreement with the value derived from the masses of the particles belonging to it (Fig. 2).

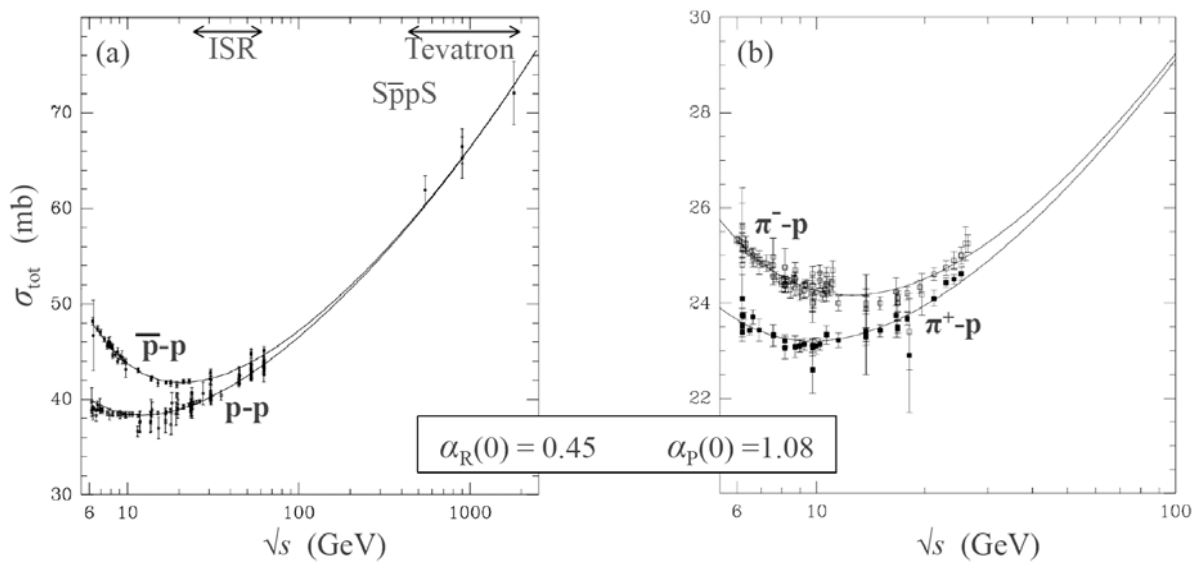

**Fig. 39:** The figure shows the fits obtained by A. Donnachie and P. Landshoff to the best-measured total cross-sections [3]

In the fit, the standard pomeron intercept is at $\alpha(0) = 1.08$ but the authors warn the reader that the exponent $\varepsilon = 0.08$ (appearing in the energy dependence $s^\varepsilon$ of the total cross-sections) is a little less than $\alpha(0) - 1$ because of multiple pomeron exchange.

They state their conclusion in the following terms: 'The fact that all cross-sections rise with energy at the same rate $s^\varepsilon$ makes it unnatural to attribute the rise to some intrinsic property of the hadrons involved. It is unhelpful to adopt a geometrical approach and to talk of hadrons becoming bigger and blacker as the energy increases. Rather the rise is a property of something that is exchanged, the pomeron, and this is why the rise is universal. … Our conclusions are in accord with the recent important results from UA8 at the CERN collider, which indicate that the pomeron does have a rather real existence: it can hit hadrons hard, break them up and knock most of their fragments sharply forward.'

This shows that twenty years ago the debate between the followers of the *s*-channel and the *t*-channel approaches to high-energy scattering phenomena was going on. And it is still alive, as indicated by a recent paper by Donnachie and Landshoff [69] who, forty years after the first ISR physics run, have analysed the data produced at the LHC by the TOTEM Collaboration [72] coming to the conclusion that their picture is still valid but a term has to be added due to the 'hard pomeron' already seen in electron–proton collisions at HERA.

## 8    Conclusions

It is often said that the ISR did not have the detectors needed to discover fundamental phenomena made accessible by its large and new energy range. This is certainly true for 'high-momentum-transfer physics', which, since the end of the 1960s, became a main focus of research, but the statement does not apply to the field that is the subject of this paper.

In fact, looking back to the results obtained at the ISR by the experiments that were programmed to study 'small-angle physics', one can safely say that the detectors were very well suited to the tasks and performed much better than foreseen.

As far as the results are concerned, in this particular corner of hadron–hadron physics, new phenomena were discovered, unexpected scaling laws were found and the first detailed studies of that elusive concept, which goes under the name 'pomeron', were performed, opening the way to phenomena that we hope will be observed at the LHC.

Moreover, some techniques and methods have had a lasting influence: all colliders had and have their Roman pots, and the different methods developed at the ISR for measuring the luminosity are still in use.

'Small-angle physics' is not very fashionable today but gave a lot of satisfaction to those who laboured around it and, in addition, has a great merit: it requires a very close collaboration among machine physicists and experimentalists, an invaluable gift that we enjoyed at the time of the ISR and for which we experimentalists are still grateful forty years later.

I am grateful to Mike Albrow, Luigi Di Lella and Kurt Hübner for suggestions and corrections to the first version of this report.